\pdfoutput=1
\documentclass[letterpaper,journal]{IEEEtran}

\usepackage{amsthm}
\usepackage{amsmath,amsfonts}
\usepackage{amssymb}
\usepackage{algorithmic}
\usepackage{algorithm}
\usepackage{array}
\usepackage[caption=false,font=footnotesize,labelfont=rm,textfont=rm]{subfig}
\usepackage{textcomp}
\usepackage{stfloats}
\usepackage{url}
\usepackage{verbatim}
\usepackage{graphicx}
\usepackage{cite}
\usepackage{bm}
\usepackage{booktabs}
\usepackage{multirow}
\usepackage{makecell}

\newtheorem*{key question}{\bf Key Question}
\newtheorem{corollary}{\bf Corollary}
\newtheorem{example}{\bf Example}
\newtheorem{definition}{\bf Definition}

\newtheorem{observation}{\bf Observation}
\newtheorem{proposition}{\bf Proposition}
\newtheorem*{mechanism}{\bf Mechanism}
\newtheorem{thm}{\bf Theorem}
\newtheorem{lem}{\bf Lemma}
\usepackage{cases}
\newtheorem{problem}{\bf Problem}
\usepackage{dsfont}
\usepackage{multirow}
\usepackage{array} 
\usepackage{modroman}
\DeclareMathOperator*{\argmax}{argmax}
\DeclareMathOperator*{\argmin}{argmin}
\usepackage{graphics}
\usepackage{epstopdf}
\usepackage{tcolorbox}

\begin{document}

\title{Mechanism Design for Federated Learning with Non-Monotonic Network Effects}

\author{Xiang~Li,~\IEEEmembership{Student Member,~IEEE,}
        Bing~Luo,~\IEEEmembership{Senior Member,~IEEE,}
        Jianwei~Huang$^*$,~\IEEEmembership{Fellow,~IEEE,}
        and~Yuan~Luo,~\IEEEmembership{Senior Member,~IEEE}
\thanks{Xiang Li is with Shenzhen Loop Area Institute, Shenzhen Institute of Artificial Intelligence and Robotics for Society, and the School of Science and Engineering, The Chinese University of Hong Kong, Shenzhen (E-mail: xiangli2@link.cuhk.edu.cn). Bing Luo is with the Data Science Research Center, Duke Kunshan University, Jiangsu (E-mail: bl291@duke.edu). Jianwei Huang is with the School of Science and Engineering, Shenzhen Institute of Artificial Intelligence and Robotics for Society, Shenzhen Key Laboratory of Crowd Intelligence Empowered Low-Carbon Energy Network, and CSIJRI Joint Research Centre on Smart Energy Storage, The Chinese University of Hong Kong, Shenzhen, Guangdong, 518172, P.R. China. He is also affiliated with the Shenzhen Loop Area Institute, Shenzhen, Guangdong, 518038, P.R. China (Corresponding Author, E-mail: jianweihuang@cuhk.edu.cn). Yuan Luo is with the School of Science and Engineering, Shenzhen Institute of Artificial Intelligence and Robotics for Society, and Shenzhen Key Laboratory of Crowd Intelligence Empowered Low-Carbon Energy Network, The Chinese University of Hong Kong, Shenzhen (E-mail: luoyuan@cuhk.edu.cn).}
\thanks{Part of the analysis appeared in ACM MobiHoc 2024 Conference \cite{10.1145/3641512.3686394}.}
}

\markboth{IEEE Transactions on Mobile Computing}%
{Shell \MakeLowercase{\textit{et al.}}: A Sample Article Using IEEEtran.cls for IEEE Journals}


\maketitle

\begin{abstract}
Mechanism design is pivotal to federated learning (FL) for maximizing social welfare by coordinating self-interested clients. Existing mechanisms, however, often overlook the network effects of client participation and the diverse model performance requirements (i.e., generalization error) across applications, leading to suboptimal incentives and social welfare, or even inapplicability in real deployments. To address this gap, we explore incentive mechanism design for FL with network effects and application-specific requirements of model performance. We develop a theoretical model to quantify the impact of network effects on heterogeneous client participation, revealing the non-monotonic nature of such effects. Based on these insights, we propose a Model Trading and Sharing (MoTS) framework, which enables clients to obtain FL models through either participation or purchase. To further address clients' strategic behaviors, we design a Social Welfare maximization with Application-aware and Network effects (SWAN) mechanism, exploiting model customer payments for incentivization. Experimental results on a hardware prototype demonstrate that our SWAN mechanism outperforms existing FL mechanisms, improving social welfare by up to $352.42\%$ and reducing extra incentive costs by $93.07\%$.
\end{abstract}


\section{Introduction}
\IEEEPARstart{F}{ederated} learning (FL)~\cite{10.1016/j.neucom.2024.128019,Zhan22,Khan20,Tu22} is an emerging machine learning (ML) paradigm that leverages massive data~\cite{Nguyen2021-zp} and resources~\cite{Guo2021-bf} across devices. With collaborative model training, FL aims to achieve the maximal social welfare (defined as \textit{the aggregate payoff for all parties involved}~\cite{Rong21}), subject to diverse application-specific requirements on model performance (i.e., generalization ability~\cite{10571602}). For example, in domains like healthcare~\cite{10288131}, beyond maximizing social welfare, the trained FL model must satisfy stringent generalization error thresholds~\cite{10.1145/3501296} to ensure reliability.\begin{figure}[!htbp]
    \centering
    \includegraphics[width=0.44\textwidth]{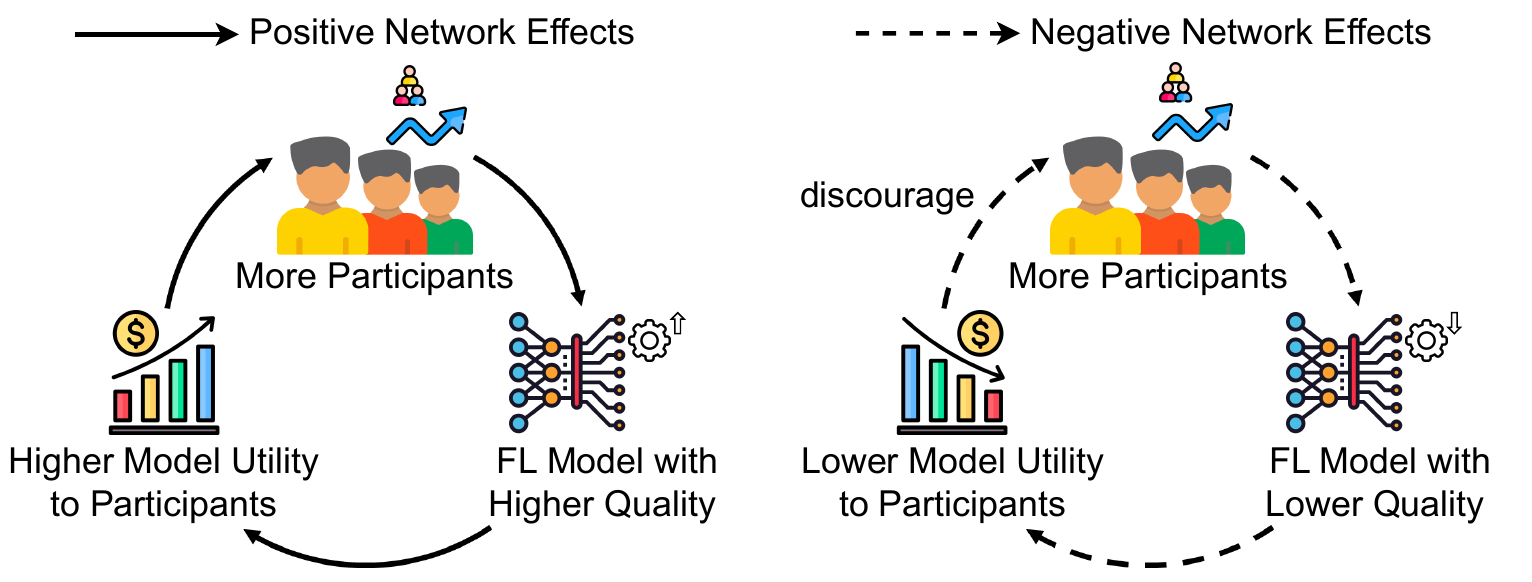}
    \vspace{-2pt}
    \caption{Network Effects of Client Participation in FL.}
    \label{NE}
\end{figure} While for areas such as recommendation systems~\cite{10262054}, this constraint is more relaxed. Realizing such application-aware social welfare maximization, however, requires active participation of heterogeneous clients, often hindered by training costs~\cite{Lee20,zhao2023truthful,karimireddy2022mechanisms}, e.g., computational costs. For existing FL frameworks (e.g., \cite{mcmahan2017communication,Tan2022,Abdul2021}), although clients can obtain the trained model by participating, the utility derived from the model may not offset their costs, resulting in reduced participation. Thus, to fully harness client collaboration, designing a mechanism that maximizes social welfare while satisfying varying performance requirements of applications is imperative for FL to succeed.

Recent studies, e.g., \cite{Thi21,Lin22,Zhan20,Jiao21,Saputra23,Luo23,10476711,9797864,9843871,9488705,11164975,10403801,9795863}, have made impressive strides in optimizing social welfare through mechanism design in FL. Nevertheless, they typically overlook the \textbf{network effects} \cite{Easley2012-kj} arising from client participation and the heterogeneous requirements of model performance across applications. Specifically, \emph{network effects\footnote{In economics, network effect is a phenomenon where product utility varies with client number, often showing monotonicity: positive (utility increases with clients, e.g., social networks) or negative (e.g., network congestion).} refer to the interdependencies between client participation and trained model utility, which vary with the network size, i.e., participant number}. These effects influence the correlations between clients' participation decisions. For instance, as more clients contribute to FL, training models with higher utility becomes more feasible, attracting more participants in a positive feedback loop. Conversely, due to data heterogeneity \cite{luoyiqian}, more participants may also yield models with lower utility, thus discouraging participation, as depicted in Figure \ref{NE}. Ignoring such network effects could lead to improper incentive mechanism design at different stages, either underpaying or overpaying clients for their participation. This misalignment of incentives is exacerbated by the diverse requirements on model performance, failing to adapt to the dynamics of clients' participation decisions. Motivated by this, we investigate the following key question in this paper:
\begin{key question}
\label{q1}
How to design an application-aware social welfare maximization mechanism for FL with network effects?
\end{key question}
\begin{figure}[!htbp]
    \centering
    \includegraphics[width=0.48\textwidth]{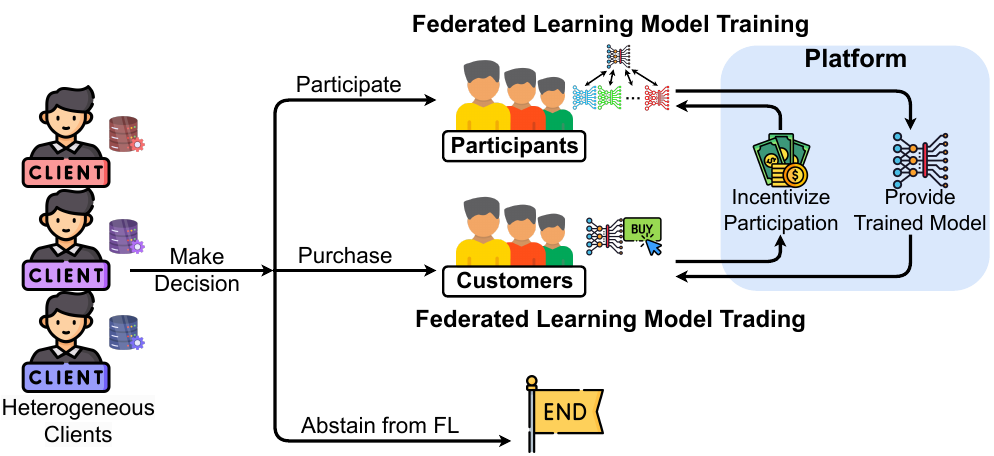}
    \vspace{6pt}
    \caption{The Model Trading and Sharing (MoTS) Framework.}
    \label{framework}
    \vspace{11pt}
\end{figure}
To address the key question above, we propose a \underline{Mo}del \underline{T}rading and \underline{S}haring (MoTS) framework, inspired by nascent ML model markets (e.g., \cite{StreamML,GravityAI,Modelplace}), to exploit FL network effects and enhance social welfare. The MoTS framework, as illustrated in Figure \ref{framework}, allows clients to obtain FL models in two ways: by participating in training as a participant, or by purchasing trained models as a model customer. With this extra purchasing option, FL models are more accessible to clients (e.g., for those with high participation costs or limited data size), and payments from model customers can \emph{trigger network effects that promote participation, increasing social welfare}.

However, self-interested clients prioritize individual payoff, which may deviate from social welfare maximization or fail to meet the performance thresholds required by specific applications. Thus, we need to design a mechanism for the MoTS framework to maximize social welfare under varying model performance requirements, which is highly challenging due to the unique characteristics of FL with network effects. In particular, the client statistical heterogeneity (i.e., data size and distribution heterogeneity \cite{Yan22}) and distributed nature of FL complicate the quantification of trained model performance. This inherent complexity necessitates sophisticated theoretical models \cite{pmlr-v119-karimireddy20a} to assess network effects, along with the expected model utility and payoff that clients derive from participation. Moreover, FL's privacy-preserving feature restricts clients from accessing each other's private information, creating an incomplete information game, while the extra purchasing options expand their decision space even further.

Against this background, we begin with linear FL models, which are insightful in understanding nonlinear scenarios, as shown in  \cite{NIPS2017_6211080f,liang2020think,Trevor22,Mei22,AKrogh_1992}. Through the linear model setting and quantifying the network effects on FL model performance, we explore the Nash equilibrium \cite{mas1995microeconomic} among multiple clients with incomplete information. Building upon these, we design the \underline{S}ocial \underline{W}elfare Maximization with \underline{A}pplication-Aware and \underline{N}etwork Effects (SWAN) mechanism. The SWAN mechanism \textit{incentivizes heterogeneous clients through payments from model customers and leverages the network effects of client participation, minimizing extra incentive costs. Across diverse FL scenarios, it can maximize social welfare while satisfying the application-specific requirements on model performance.}

The key results and contributions of the paper are as follows:
\begin{itemize}
    \item \emph{New Problem Formulation:} 
    This paper presents the first application-aware social welfare maximization framework for federated learning that explicitly incorporates non-monotonic network effects of client participation, addressing a critical gap in existing FL mechanism design.
    \item \emph{Fundamental Theoretical Insight:} 
    By rigorously analyzing generalization error under heterogeneous data sizes and distributions, the paper reveals a counter-intuitive, non-monotonic structure of FL network effects, showing that client participation can transition between positive and negative impacts on model performance depending on data heterogeneity.
    \item \emph{Core Technical Solution:} The paper proposes the Model Trading and Sharing (MoTS) framework and designs the SWAN mechanism, which internalizes network effects and incomplete-information strategic behavior to jointly optimize pricing and incentives, achieving application-specific performance guarantees while minimizing (and sometimes eliminating) incentive costs.
    \item \emph{Practical Validation and Impact:} Extensive experiments on a real FL hardware prototype validate the theoretical analysis and demonstrate substantial gains over existing FL mechanisms, with up to $352.42\%$ improvement in social welfare and up to $93.07\%$ reduction in incentive costs, while supporting diverse application requirements.
\end{itemize}

The remainder of this paper is organized as follows. Section~\ref{S1} reviews related works of mechanism design and network effects in FL. Section~\ref{S2} introduces the system model and problem formulation. Section~\ref{S3} presents our proposed MoTS framework and, drawing on the analysis of network effects, develops the corresponding SWAN mechanism. Section~\ref{S4} provides the experimental evaluation on an FL hardware prototype. We conclude this paper in Section~\ref{S5}. Due to the page limit, we leave the detailed proof of our results and additional empirical experiments in the online appendix~\cite{Online_appendix}.

\section{Related Works}\label{S1}
In this section, we review both the mechanism design for social welfare maximization and network effects in FL, identifying critical gaps that motivate our contributions.

\textbf{Mechanism Design for FL Social Welfare.} The mechanism design for FL social welfare optimization has made considerable progress (e.g.,~\cite{Thi21,Lin22,Zhan20,Jiao21,Saputra23,Luo23,10476711,9797864,9843871,9488705,11164975,10403801,9795863}), as summarized in recent surveys~\cite{Zhan22,Tu22,Khan20,Nguyen2021-zp,Guo2021-bf,Rong21}. Specifically, Lee \emph{et al.} in~\cite{Lee20} formulated the interactions among FL participants as a Stackelberg game, aiming to maximize social welfare. Zhan \emph{et al.} in~\cite{Zhan20} introduced a learning-based algorithm to evaluate ambiguous client contributions, which improves the aggregate payoff for all participants. Jiao \emph{et al.} in~\cite{Jiao21} proposed an auction-based market model for incentivizing clients to participate in FL, thereby effectively maximizing the social welfare. Saputra \emph{et al.} in~\cite{Saputra23} developed a contract-based policy to increase the social welfare of the FL framework. 

However, these works typically impose no constraints on the trained model performance and overlook network effects. As a result, they may fail to satisfy the performance requirements of specific applications and create inefficient incentives: underpaying or overpaying clients for their participation. In contrast, this paper theoretically analyzes the network effects of client participation, which is relatively underexplored and more complex given the unique attributes of FL. We develop a SWAN mechanism within the MoTS framework that leverages network effects to maximize social welfare. Our mechanism can meet various application-specific requirements on FL model performance (i.e., generalization ability) while minimizing the additional incentive costs.

\textbf{Network Effects in FL.} The only work that studies FL network effects is~\cite{hu2023} by Hu \emph{et al.}, which modeled client behavior as a network effects game to analyze the participation equilibrium. However, their analysis assumed uniform client expectations and monotonic performance improvement with participation, without fully considering client heterogeneity and the impact on social welfare. In practice, clients often lack information about others' decisions, and data statistical heterogeneity could cause complex dependencies between participation and model performance. This incomplete information and heterogeneity could lead to their mechanism failing to incentivize diverse clients towards the desired equilibrium. In contrast, this paper designs a mechanism to incentivize the participation of heterogeneous clients with incomplete information, aiming to maximize social welfare under varying model performance requirements for FL applications.

\section{System Model and Problem Formulation}
\label{S2}
This section establishes our Model Trading and Sharing (MoTS) framework for FL that enables both participation and purchasing options, and formulates the mechanism design problem of application-aware social welfare maximization. We begin with FL fundamentals, characterize client heterogeneity and network effects by generalization error analysis, then present our MoTS framework and optimization problem.

\subsection{Federated Learning with Heterogeneous Clients}
\subsubsection{FL Training Process}
We consider a set of $N$ clients interested in FL model, denoted by $\mathcal{N}=\{1, 2, \cdots, N\}$. Let $\mathcal{K}\subseteq\mathcal{N}$ be the set of clients that actually participate in the training process. Each participant $k\in\mathcal{K}$ possesses a private dataset $\mathcal{D}_k$ of size $D_k$ and trains a local model on it. The local training objective of participant $k$ is to find the optimal model parameters that minimize the empirical loss function $F_k(\boldsymbol{w})$ over all data samples $d\in\mathcal{D}_k$:
\begin{equation}
    F_k(\boldsymbol{w})=\frac{1}{D_k}\sum_{d\in\mathcal{D}_k} f(\boldsymbol{w};d),
\end{equation}
where $f(\cdot)$ is the per-sample loss function.

With locally trained models, FL solves the global loss optimization problem over all participants' data, i.e., to learn the optimal global model parameters $\boldsymbol{w}^*$ by minimizing the weighted average of participants' local losses:
\begin{equation}
    \boldsymbol{w}^*=\arg\min_{\boldsymbol{w}} F(\boldsymbol{w})=\arg\min_{\boldsymbol{w}}\sum_{k\in\mathcal{K}}\frac{D_k}{\sum_{k\in\mathcal{K}}D_k}F_k(\boldsymbol{w}).
\end{equation}
\subsubsection{Client Heterogeneity} We model two forms of statistical heterogeneity across the client set $\mathcal{N}$:
\begin{itemize}
    \item \textit{Data distribution heterogeneity:} Following~\cite{Trevor22,NIPS2017_6211080f,liang2020think}, local features follow independent Gaussian distributions with diagonal covariance $\sigma^2\in\mathbb{R}^d$, where $d$ is the feature dimension and a larger $\sigma^2$ indicates greater divergence among clients’ data (i.e., higher client variance). For naturally occurring data variance, we capture it as Gaussian noise $\delta\sim\mathcal{N}(0,\gamma^2)$ on the training targets.
    \item \textit{Data size heterogeneity}: We categorize clients into $I$ types $\mathcal{I}=\{1,2,\ldots,I\}$ by their data size $D_i$. Let $\mathcal{N}_i$ denote the set of $N_i$ type-$i$ clients, where $\bigcup_{i\in\mathcal{I}}\mathcal{N}_i=\mathcal{N}$ and $\sum_{i\in\mathcal{I}}N_i=N$. Without loss of generality, we assume $D_i\leq D_j$, if $i\leq j$, for any $i,j\in\mathcal{I}$.
\end{itemize}

\subsubsection{Participation Cost and Model Utility} Type-$i$ clients incur participation cost $C_i$ during FL model training, where $C_i\leq C_j$ for $i\leq j$, reflecting resource consumption that correlates\footnote{This positive correlation is only to align with real-world situations, which will not affect our theoretical analysis and mechanism design later.} with data size, such as data labeling, computation, and communication cost~\cite{Lee20,zhao2023truthful,karimireddy2022mechanisms}. The client utility derived from the model depends on its generalization performance~\cite{10571602}, quantified by generalization error $\varepsilon$, which measures how well the trained FL model performs on unseen data. We denote $U(\varepsilon)\geq 0$ as clients' model utility function~\cite{Chen2019-js,10.14778/3447689.3447700,10.1145/3328526.3329589}, where $U'(\varepsilon)\leq0$ indicates that a lower generalization error yields higher utility. Motivated by practical observations~\cite{Fernandez2020-cx,10.1145/3299869.3300078,Cong2022-pp}, we assume that $(\varepsilon-\sigma^2)U''(\varepsilon)+2U'(\varepsilon)\geq 0$ holds for $\varepsilon\neq\sigma^2$, which characterizes the increasing marginal utility~\cite{hestness2017deep} as $\varepsilon$ approaches client variance threshold $\sigma^2$.

\subsubsection{Generalization Error and Network Effects}
Extending the analytical framework in~\cite{liang2020think} to heterogeneous settings, we derive generalization error\footnote{Due to page limit, we give the derivation of~\eqref{E7} in online appendix~\cite{Online_appendix}.} $\varepsilon$ of the trained FL model as:
\begin{equation}
    \label{E7}
        \varepsilon=\frac{d\gamma^2}{K^2}\sum_{i\in\mathcal{I}}\frac{K_i}{D_i}+\frac{K-1}{K}\sigma^2,
\end{equation}
where $K_i$ denotes the number of type-$i$ participants and $K=\sum_{i\in\mathcal{I}}K_i$. This reveals two competing effects of participation: data variance $\gamma^2$ induces error inversely related to participant number and the harmonic mean of their data size, while client variance $\sigma^2$ increases error with more participants, leading to non-monotonic network effects analyzed in Section~\ref{S4.1}.
\subsection{Model Trading and Sharing (MoTS) Framework} 
To exploit network effects while addressing participation disincentives, we propose MoTS framework, enabling clients to either participate in training or purchase the trained model.

\subsubsection{Client Strategies} In the MoTS framework, each client $n\in\mathcal{N}_i$ chooses strategy\footnote{To deliver clear insights, we focus on full participation behaviors of clients rather than partial ones (i.e., randomly joining some rounds of FL training), where the latter will yield similar analytical results, as studied in~\cite{wang2024a,yang2021achievinglinearspeeduppartial,10292582}.} $s_{n}\in\{A,J,B\}$:
\begin{subnumcases}{\label{E3}s_{n}=}
    A, &\textrm{abstain from obtaining the FL model},\\
    J, &\textrm{join FL training to obtain the model},\\
    B, &\textrm{buy the FL model}.
\end{subnumcases}

Given client strategies, we define $K_i=\sum_{n\in\mathcal{N}_i}\mathds{1}_{\{s_{n}=J\}}$ and $B_i=\sum_{n\in\mathcal{N}_i}\mathds{1}_{\{s_{n}=B\}}$ as the numbers of type-$i$ participants and model buyers, respectively. The indicator function $\mathds{1}_{\{s_{n}=J\}}$ means that client $n$ chooses participation in FL model training, whereas $\mathds{1}_{\{s_{n}=B\}}$ indicates the client's choice to purchase the trained model.

\subsubsection{Platform Mechanism}\label{plat} The MoTS platform, as illustrated in Figure~\ref{framework}, coordinates the FL model training and trading process with model price $p$ and type-$i$ participation reward $r_i$. To focus on social welfare maximization, we consider a non-profit\footnote{For the platform incurring operational costs, the analysis is the same, but incorporates these costs into the overall evaluation of social welfare.} platform, leveraging all buyer payments to incentivize client participation and model improvements, such that:
\begin{equation}
    \sum_{i\in\mathcal{I}}B_i\cdot p\leq\sum_{i\in\mathcal{I}}K_i\cdot r_i.
\end{equation}

The difference $\sum_{i\in\mathcal{I}}(B_i\cdot p - K_i\cdot r_i) \leq 0$ captures the platform's additional incentive costs, where $r_i$ can be negative to penalize low-contributing participants.

\subsubsection{Client Payoff}
For type-$i$ client $n\in\mathcal{N}_i$, the payoff $\pi_n$ depends on all clients' strategies $\boldsymbol{s}$:
\begin{subnumcases}{\label{E6}\pi_{n}(\boldsymbol{s})=}
    0,&\textrm{if $s_{n}=A$},\label{6a}\\
    U(\varepsilon)-C_i+r_{i},&\textrm{if $s_{n}=J$},\label{6b}\\
    U(\varepsilon)-p,&\textrm{if $s_{n}=B$}.\label{6c}
\end{subnumcases}

From~\eqref{6a}, if client $n$ abstains from acquiring the trained global FL model, the payoff $\pi_n$ is zero. By participating in the training process, the type-$i$ client $n\in\mathcal{N}_i$ receives model utility $U(\varepsilon)$ and participation reward $r_{i}$, while incurring participation cost $C_i$, as shown in~\eqref{6b}. If client $n$ purchases the FL model directly, the payoff $\pi_n$ is the difference between model utility $U(\varepsilon)$ and model price $p$, as given in~\eqref{6c}.
\subsection{Application-Aware Social Welfare Maximization}\label{S3D}
With the proposed MoTS framework, the platform aims to design a mechanism that maximizes social welfare while satisfying application-specific FL performance requirements.
\subsubsection{Social Welfare Definition}
In the MoTS framework, the social welfare is the sum of each client's payoff $\pi_{n}$ in~\eqref{E6} and the platform's payoff $\sum_{i\in\mathcal{I}}(B_i\cdot p-K_i \cdot r_i)$.
\begin{definition}[Social State of FL~\cite{Easley2012-kj}]
    \label{D1} A social state $\{\boldsymbol{K},\boldsymbol{B}\}=\{K_i, B_i\}_{i \in\mathcal{I}}$ specifies the number of type-$i$ clients choosing strategy $J$ and $B$ respectively, for all $i\in\mathcal{I}$.
\end{definition}
Given social state $\{\boldsymbol{K},\boldsymbol{B}\}=\{K_i, B_i\}_{i \in\mathcal{I}}$ in Definition~\ref{D1}, the social welfare $W_{\textrm{MoTS}}(\boldsymbol{K},\boldsymbol{B})$ is expressed as follows:
\begin{equation}\label{E5}
    W_{\textrm{MoTS}}(\boldsymbol{K},\boldsymbol{B})=\sum_{i\in\mathcal{I}}[K_i\cdot(U(\varepsilon)-C_i)+B_i\cdot U(\varepsilon)].
\end{equation}
\subsubsection{Optimization Problem}
We formulate the platform's mechanism design problem for application-aware social welfare maximization as Problem~\ref{Pb1}.
\begin{figure}[!htbp]\begin{tcolorbox}
\begin{problem}[Application-Aware Social Welfare Maximization]\label{Pb1}
    \begin{align}
    \max_{\{p,r_i\}_{i\in\mathcal{I}}}&\sum_{i\in\mathcal{I}}\big[K^*_i\cdot(U(\varepsilon)-C_i)+B^*_i\cdot U(\varepsilon)\big],\label{Pb1obj}\\
    s.t.\textrm{ }&\varepsilon(\boldsymbol{K^*})\leq\varepsilon_{\emph{req}},\textrm{ } \varepsilon_{\emph{req}}\in[\varepsilon_{\min},+\infty),\label{Pb1perf}\\
    &\{p,r_i\}_{i\in\mathcal{I}}=\argmin_{\{\hat p,\hat r_i\}_{i\in\mathcal{I}}}\left[\sum_{i\in\mathcal{I}}K^*_i \hat r_i-B^*_i  \hat p\right]^+,\label{Pb1cost}\\
    &s^*_{n}=\argmax_{s_{n}\in\{A,J,B\}} \pi_{n}\left(s_{n},\boldsymbol{s}^*_{-n}\right),\textrm{ }\forall n\in\mathcal{N},\label{Pb1client}\\
    &K^*_i=\sum_{n\in\mathcal{N}_i}\mathds{1}_{\{s^*_{n}=J\}},\textrm{ }\forall i\in\mathcal{I},\\
    &B^*_i=\sum_{n\in\mathcal{N}_i}\mathds{1}_{\{s^*_{n}=B\}},\textrm{ }\forall i\in\mathcal{I},
\end{align}
where $[x]^+ \triangleq \max\{0,x\}$ and $\varepsilon$ follows~\eqref{E7}.
\end{problem}
\end{tcolorbox}
\end{figure}

In Problem~\ref{Pb1}, constraint~\eqref{Pb1perf} ensures application performance requirements with $\varepsilon_{\min}$ as the minimum achievable generalization error. Constraint~\eqref{Pb1cost} minimizes additional incentive costs and~\eqref{Pb1client} captures client optimality, guaranteeing the existence and uniqueness of the client-side Nash equilibrium. The following section designs a Social Welfare Maximization with Application-Aware and Network Effects (SWAN) mechanism to solve this challenging optimization problem.

\section{Mechanism Design with Network Effects}
\label{S3}
This section develops a comprehensive framework for understanding and leveraging network effects in FL. We first establish the theoretical foundations of FL network effects, revealing their non-monotonic nature and dependence on client heterogeneity. Based on this analysis, we derive optimal social states that maximize social welfare under application-specific performance constraints. Finally, we propose our Social Welfare Maximization with Application-Aware and Network Effects (SWAN) mechanism that aligns individual client incentives with social optimality while minimizing platform costs.

\subsection{Network Effects of Client Participation}\label{S4.1}
FL enables clients to collaboratively train and share a global model with the aim of generating accurate outputs for a particular task. From the generalization error $\varepsilon$ in~\eqref{E7}, we begin by analyzing the impact of client collaboration in FL and then exploring the network effects of client participation.

\textbf{Conditions for Error-Reducing Collaboration.} We first identify the conditions under which client collaboration improves model performance, laying the foundations for understanding FL network effects, as formalized in Corollary~\ref{C1}.
\begin{corollary}
\label{C1}
    Consider two disjoint participant coalitions $\mathcal{K}_a\neq\emptyset$ and $\mathcal{K}_b\neq\emptyset$, where $H_a$ and $H_b$ denote the harmonic means of participants' data sizes in $\mathcal{K}_a$ and $\mathcal{K}_b$, respectively, and $H_a\leq H_b$. The generalization error $\varepsilon_{a+b}$ of the FL model trained by $\mathcal{K}_a\cup\mathcal{K}_b$ is lower than that of the model trained by either coalition, i.e., $\varepsilon_{a+b}<\max\{\varepsilon_a,\varepsilon_b\}$, if and only if
    \begin{align}
        \frac{\sigma^2}{\gamma^2}<\frac{H_bK_b+K_a(2H_b-H_a)}{H_aH_b(K_a+K_b)/d}.
    \end{align}
\end{corollary}

We give the proof of Corollary~\ref{C1} in online appendix~\cite{Online_appendix}. Corollary~\ref{C1} implies that client collaboration in FL can benefit individual participants (or coalitions of multiple participants) by enabling them to obtain a model with lower generalization error $\varepsilon$ compared to training alone. Nevertheless, collaboration is not always beneficial to all participants, especially when client variance $\sigma^2$ outweighs data variance $\gamma^2$ by a certain threshold (related to the harmonic mean of their data sizes, i.e., precision weight~\cite{hu2023generalization}). In such cases, a client's participation could increase the generalization error of the FL model trained by existing participants. This decline in model performance reduces model utility for existing participants, exhibiting negative network effects of client participation in FL.

\textbf{Network Effects Definition and Characterization.} To thoroughly investigate the varying network effects in FL, we quantify them as the change in generalization error $\varepsilon$ of the trained model caused by client participation, defined in Definition~\ref{NECP}. Then, we determine the boundary between positive and negative FL network effects in Theorem~\ref{T1}.

\begin{definition}[Network Effects~\cite{Easley2012-kj} of Client Participation]
    \label{NECP} Given any non-empty coalition of existing participants $\mathcal{K}$, the participation of a client $n$ results in network effects, quantified as the difference between generalization error $\varepsilon_{\mathcal{K}}$ of the FL model trained by $\mathcal{K}$ and the generalization error $\varepsilon_{\mathcal{K}\cup\{n\}}$ of the model trained by $\mathcal{K}\cup\{n\}$, i.e., $\varepsilon_{\mathcal{K}}-\varepsilon_{\mathcal{K}\cup\{n\}}$.
\end{definition}

\begin{thm}
    \label{T1}
    Consider any non-empty coalition of existing participants $\mathcal{K}$, the participation of a client $n$ with data size $D$ will bring non-negative network effects, i.e., $\varepsilon_{\mathcal{K}}-\varepsilon_{\mathcal{K}\cup\{n\}}\geq 0$, if and only if ${1}/{D}$ is no more than $\eta$ in~\eqref{delta}, i.e., ${1}/{D}\leq \eta$,
    \begin{equation}\label{delta}
        \eta=\frac{(2K+1)\sum_{i\in\mathcal{I}}{K_i}/{D_i}}{K^2}-\frac{(K+1)\sigma^2}{d\gamma^2K},
    \end{equation}
where $K_i$ is the number of type-$i$ participants in $\mathcal{K}$, for any $i\in\mathcal{I}$, and $K=\sum_{i\in\mathcal{I}}K_i$.
\end{thm}

We give the proof of Theorem~\ref{T1} in online appendix~\cite{Online_appendix}.
With $\eta$ in~\eqref{delta}, Theorem~\ref{T1} identifies two factors that influence the network effects of client participation: the data size of the new participant against that of existing participants, and the ratio of client heterogeneity $\sigma^2$ to data heterogeneity $\gamma^2$. 
\begin{itemize}
    \item If the ratio of client heterogeneity to data heterogeneity ${\sigma^2}/{\gamma^2}$ exceeds the threshold $\frac{(2K+1)\sum_{i\in\mathcal{I}}{K_i}/{D_i}}{K(K+1)/d}$, client participation will increase the trained model's generalization error $\varepsilon$, resulting in negative network effects. 
    \item Conversely, when ${\sigma^2}/{\gamma^2}$ is below this threshold, a new participant requires only one-third of the harmonic mean of existing participants' data sizes to positively impact the current model (i.e., reducing error $\varepsilon$), as deduced by the upper bound $({3\sum_{i\in\mathcal{I}}K_iD_i})/{K}$ on $\eta$ in~\eqref{delta}. 
\end{itemize}

However, the composition of participants evolves with the participation of new clients, which in turn affects the specific data size threshold required for future new clients to ensure non-negative network effects from their participation.

\textbf{Illustrative Example: Two Client Types.} Given the above discussions, heterogeneous participants jointly influence both the performance of the FL model and the network effects of client participation. To demonstrate how the model performance and network effects vary with the number of participants of different types, we first take $I=2$ types of clients, each possessing homogeneous datasets (i.e., $\sigma^2=0$), as an illustrative example, shown in Example~\ref{E1}.
\begin{example}[Two Types of Clients]
\label{E1}
    Consider $I=2$ types of clients with $\sigma^2=0$ in FL, and a coalition of participants $\mathcal{K}$. As the number $K_1$ of type-$1$ participants in $\mathcal{K}$ increases, the generalization error $\varepsilon$ of the trained global FL model initially increases, but eventually decreases once $K_1\geq\lceil{({-2K_2D_1-D_2+\sqrt{4K_2^2(D_2-D_1)^2+D_2^2}})/{2D_2}}\rceil$. In contrast, the generalization error $\varepsilon$ consistently decreases as the number $K_2$ of type-$2$ participants in $\mathcal{K}$ grows.
\end{example}
Example~\ref{E1} provides an interesting insight under the independent and identically distributed (i.i.d.) data distribution: when the number of participants with small data size falls below a certain threshold (see type-$1$ clients in Example~\ref{E1}), their participation incurs negative network effects; however, once the number of these participants exceeds this threshold, their participation yields positive network effects. In contrast, the participation of clients with the largest data size consistently brings positive network effects, regardless of their number.

\textbf{General Network Effects Analysis.} After introducing the case with two client types and i.i.d. data distribution, we extend our analysis to more general settings involving $I$ client types and non-i.i.d. data distribution, as characterized in Theorem~\ref{T2}.

\begin{thm}
\label{T2}
    Consider a coalition of existing participants $\mathcal{K}\neq\emptyset$, where $K_i$ is the number of type-$i$ participants in $\mathcal{K}$, for any $i\in\mathcal{I}$, and $K=\sum_{i\in\mathcal{I}}K_i$. Then, for clients with type $j\in\mathcal{I}$,
    \begin{itemize}
    \item if ${1}/{D_j}<{\sigma^2}/{(d\gamma^2)}$ and ${1}/{D_j}\leq\eta$, network effects of client participation are initially non-negative, but eventually become negative as more type-$j$ clients participate;
    \item if ${\sigma^2}/{(d\gamma^2)}\leq{1}/{D_j}\leq\eta$, network effects of client participation are always non-negative;
    \item if $\eta<{1}/{D_j}\leq{\sigma^2}/{(d\gamma^2)}$, network effects of client participation are always negative;
    \item if $\max\{{\sigma^2}/{(d\gamma^2)},\eta\}<{1}/{D_j}$, network effects of client participation are initially negative, but eventually become positive as more type-$j$ clients participate.
\end{itemize}
\end{thm}
\begin{figure}[ht]
    \centering
    \includegraphics[width=0.4\textwidth]{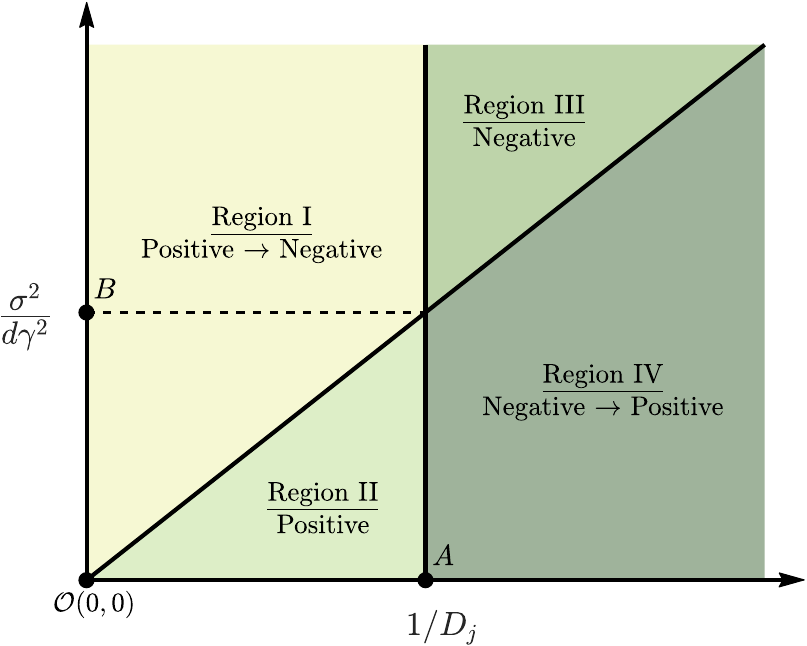}\vspace{2pt}
    \small\boxed{
\begin{array}{ll}
\text{A}=({((2K+1)\sum_{i\in\mathcal{I}}{K_i}/{D_i}})/{K^2}-{(K+1)\sigma^2}/{d\gamma^2K},0).\\
\text{B}=(0,{((2K+1)\sum_{i\in\mathcal{I}}{K_i}/{D_i}})/{K^2}-{(K+1)\sigma^2}/{d\gamma^2K}).
\end{array}}\vspace{1pt}
    \caption{Network Effects of Type-$j$ Client Participation.}
    \label{Region}
\end{figure}
We give the proof of Theorem~\ref{T2} in online appendix~\cite{Online_appendix}. From Theorem~\ref{T2}, the network effects of type-$j$ client participation not only depend on heterogeneity ratio $\sigma^2/\gamma^2$ and data size $D_j$, but may also vary with the number of type-$j$ participants $K_j$, for any $j\in\mathcal{I}$. Specifically, we can divide the network effects of client participation into four regions, as depicted in Figure~\ref{Region}. In Region~\nbRoman{1}, the trained FL model performance (and utility) first increases and then decreases as the number of same-type participants grows. In contrast, for Region~\nbRoman{2}, the network effects are always positive, leading to a monotonic reduction in generalization error $\varepsilon$. Region~\nbRoman{3} and Region~\nbRoman{4}, however, exhibit opposite trends to Region~\nbRoman{2} and Region~\nbRoman{1}, respectively. With this four-region structure, given any participant coalition $\mathcal{K}\neq\emptyset$ and heterogeneity ratio $\sigma^2/\gamma^2$, we can determine the network effects of each client type and, when applicable, the exact threshold at which network effects reverse. For instance, when $\sigma^2=0$ (representing an i.i.d. data distribution, indicated by line $OA$ in Figure~\ref{Region}), the network effects of client participation will correspond to Regions~\nbRoman{2} or~\nbRoman{4}, aligning with our observation in Example~\ref{E1}.

So far, through the theoretical analysis of FL model performance, we have investigated network effects arising from the participation of heterogeneous clients. Next, we explore the social states for application-aware social welfare maximization under the MoTS framework, aiming to develop a corresponding mechanism that leverages network effects.

\subsection{The Social States for Application-Aware Social Welfare Maximization under the MoTS Framework}
The social states $\{\boldsymbol{K},\boldsymbol{B}\}=\{K_i, B_i\}_{i \in\mathcal{I}}$, defined in Definition~\ref{D1}, determine FL social welfare and trained model performance (quantified by generalization error $\varepsilon$) within the MoTS framework. To achieve application-aware social welfare maximization, we investigate the optimal social states $\{\boldsymbol{K}^*,\boldsymbol{B}^*\}$ that maximize social welfare under application-specific requirement $\varepsilon_{\textrm{req}}$ on generalization error $\varepsilon$, i.e., $\{\boldsymbol{K}^*,\boldsymbol{B}^*\}=\argmax_{\boldsymbol{K},\boldsymbol{B}} W_\textrm{MoTS}(\boldsymbol{K},\boldsymbol{B})$, subject to $\varepsilon(\boldsymbol{K}^*)\leq\varepsilon_{\textrm{req}}$. This identifies strategies that the mechanism for solving Problem~\ref{Pb1} seeks to incentivize heterogeneous clients to adopt, forming the foundation for the design of our SWAN mechanism.

\textbf{Social Welfare Maximization Problem Reformulation.} For analytical tractability, we first derive the structural property of social states for application-aware social welfare maximization in Lemma~\ref{L1}, which allows us to later reformulate $W_{\textrm{MoTS}}(\boldsymbol{K}^*,\boldsymbol{B}^*)$ into an equivalent yet simpler form.
\begin{lem}
\label{L1}
    In optimal social states $\{\boldsymbol{K}^*,\boldsymbol{B}^*\}$ for application-aware social welfare maximization, where $\{\boldsymbol{K}^*,\boldsymbol{B}^*\}=\argmax_{\boldsymbol{K},\boldsymbol{B}} W_\emph{MoTS}(\boldsymbol{K},\boldsymbol{B})$, subject to $\varepsilon(\boldsymbol{K}^*)\leq\varepsilon_{\emph{req}}$: all clients will obtain the FL model, i.e., $\sum_{i\in\mathcal{I}}(K^*_i+B^*_i)=N$. 
\end{lem}
We give the proof of Lemma~\ref{L1} in online appendix~\cite{Online_appendix}. Lemma~\ref{L1} establishes a necessary condition of social states for maximizing social welfare while satisfying generalization error $\varepsilon_{\textrm{req}}$ required by FL applications. Specifically, for any feasible error constraint $\varepsilon_{\textrm{req}}\in[\varepsilon_\textrm{min},+\infty)$, all the clients will obtain the trained global FL model in optimal social states $\{\boldsymbol{K}^*,\boldsymbol{B}^*\}$, either by participating in training or purchasing it directly, i.e., $\sum_{i\in\mathcal{I}}(K^*_i+B^*_i)=N$. This widespread model dissemination enables each client to benefit from the trained model, thereby achieving application-aware social welfare maximization.

Based on Lemma~\ref{L1} and generalization error $\varepsilon$ given in~\eqref{E7}, we formulate Problem~\ref{Pb2} to determine the optimal social states $\{\boldsymbol{K}^*,\boldsymbol{B}^*\}$ under the MoTS framework. As a transformation of Problem~\ref{Pb1}, Problem~\ref{Pb2} preserves the objective function $W_{\textrm{MoTS}}(\boldsymbol{K},\boldsymbol{B})$ in~\eqref{Pb1obj}, offering crucial insights into the incentive structure of mechanisms for addressing Problem~\ref{Pb1}.
\begin{figure}[!htbp]
\begin{tcolorbox}
\begin{problem}[The Social States for Application-Aware Social Welfare Maximization of Problem \ref{Pb1}]\label{Pb2}
    \begin{align*}
    \max\quad &\sum_{i\in\mathcal{I}} \big[N_i\cdot U(\varepsilon)-K_i\cdot C_i\big],\\
    s.t.\quad&\varepsilon(\boldsymbol{K})\leq\varepsilon_{\emph{req}},\textrm{ } \varepsilon_{\emph{req}}\in[\varepsilon_\emph{min},+\infty),\\
    &\varepsilon=\frac{d\gamma^2}{(\sum_{i\in\mathcal{I}}K_i)^2}\sum_{i\in\mathcal{I}}\frac{K_i}{D_i}+\frac{\sum_{i\in\mathcal{I}}K_i-1}{\sum_{i\in\mathcal{I}}K_i}\sigma^2,\\
    & K_i+ B_i= N_i,\quad\forall i\in\mathcal{I},\\
    var.\quad &K_i\in\mathbb{N},\textrm{ } B_i\in\mathbb{N},\quad\forall i\in\mathcal{I}.
\end{align*}
\end{problem}
\end{tcolorbox}
\end{figure}

\textbf{Optimal Social States Characterization.} To solve Problem~\ref{Pb2}, we partition the type set $\mathcal{I}$ of heterogeneous clients into two subsets: $\mathcal{I}_L=\{i\in\mathcal{I}:D_i\leq d\gamma^2/\sigma^2\}$ (low type index) and $\mathcal{I}_H=\{i\in\mathcal{I}:D_i> d\gamma^2/\sigma^2\}$ (high type index). This partition is based on two key factors that affect the network effects of client participation, i.e., client data size $D_i$ and heterogeneity ratio $\sigma^2/\gamma^2$, as elaborated in Theorem \ref{T2}. By analyzing these two sets of clients separately, we determine the participation pattern of clients in optimal social states $\{\boldsymbol{K}^*,\boldsymbol{B}^*\}$, as stated in Proposition~\ref{P1}. 

\begin{proposition}\label{P1}
Consider any non-empty set of client types $\mathcal{I}=\mathcal{I}_{L}\cup\mathcal{I}_{H}$, where $\mathcal{I}_L=\{i\in\mathcal{I}:D_i\leq d\gamma^2/\sigma^2\}$, and $\mathcal{I}_H=\{i\in\mathcal{I}:D_i> d\gamma^2/\sigma^2\}$. In the optimal social states $\{\boldsymbol{K}^*,\boldsymbol{B}^*\}$ for application-aware social welfare maximization,
\begin{itemize}
    \item if $\mathcal{I}_H=\emptyset$, then $K_i^*\in\{0,N_i\}$, $\forall i\in\mathcal{I}_L$;
    \item if $\mathcal{I}_H\neq\emptyset$, then $K^*_i=0$, $\forall i\in\mathcal{I}_L$, and $K_j^*\in\{0,N_j\}\cup\{K_j\in\mathbb{N}:\Delta L(K_j)\leq0\leq{\Delta L(K_j-1)}\}$, $\forall j\in\mathcal{I}_H$, where $\Delta L(x)=L(x+1)-L(x)$, $L(K_j)=W_{\emph{MoTS}}(K_j,N_j-K_j)+\lambda[(\varepsilon(K_j)-\sigma^2)^{-1}-(\varepsilon_{\emph{req}}-\sigma^2)^{-1}]$ and $\lambda$ is the Lagrange multiplier associated with the error constraint $\varepsilon(\boldsymbol{K}) \leq \varepsilon_{\emph{req}}$, $\varepsilon_{\emph{req}}\in[\varepsilon_\emph{min},+\infty)$.
\end{itemize}
\end{proposition}
We give the proof of Proposition~\ref{P1} in online appendix~\cite{Online_appendix}. Proposition~\ref{P1} reveals that the optimal social states $\{\boldsymbol{K}^*,\boldsymbol{B}^*\}$ generally follow an ``all-or-none'' participation pattern. However, for clients with type $j \in \mathcal{I}_H$, partial participation in training may also achieve optimality. The underlying rationale is consistent with Theorem~\ref{T2}. According to Figure~\ref{Region}, clients of type $i\in\mathcal{I}_L$ fall into Region~\nbRoman{2} and Region~\nbRoman{4}, while clients of type $j\in\mathcal{I}_H$ occupy Region~\nbRoman{1} and Region~\nbRoman{3}. Consequently, in the optimal social states $\{\boldsymbol{K}^*,\boldsymbol{B}^*\}$, type-$i$ clients either all participate to exploit the positive network effects, or all abstain from training to avoid incurring negative ones. In contrast, clients of type $j\in\mathcal{I}_H$ exhibit an additional optimum for $\{\boldsymbol{K}^*,\boldsymbol{B}^*\}$: only part of the clients participate in training, sufficient to reap positive network effects but not so much as to trigger negative ones (i.e., reaching the threshold of participant number $K^*_j=\{K_j\in\mathbb{N}:\Delta L(K_j)\leq0\leq{\Delta L(K_j-1)}\}$ at which network effects turn from positive to negative). 

To summarize, in scenarios with low data distribution heterogeneity (where $\sigma^2\leq d\gamma^2/D_I$, yielding $\mathcal{I}_H=\emptyset$), achieving application-aware social welfare maximization requires clients of the same type to make consistent participation decisions, i.e., $K_i^*\in\{0,N_i\}$, for any type $i\in\mathcal{I}_L$. Conversely, with high data distribution heterogeneity (where $\sigma^2>d\gamma^2/D_I$, yielding $\mathcal{I}_H\neq\emptyset$), the optimal social states $\{\boldsymbol{K}^*,\boldsymbol{B}^*\}$ exclude participants of type $i\in\mathcal{I}_L$, and only clients of type $j\in\mathcal{I}_H$ potentially contribute in FL model training.

\textbf{Welfare-Performance Alignment Conditions.} After deriving the optimal social states $\{\boldsymbol{K}^*,\boldsymbol{B}^*\}$, we investigate the consistency between maximizing social welfare $W_\textrm{MoTS}(\boldsymbol{K},\boldsymbol{B})$ and minimizing generalization error $\varepsilon(\boldsymbol{K})$. Proposition~\ref{P2} establishes sufficient conditions under which these two objectives align, thereby simplifying Problem~\ref{Pb1}. With the conditions of Proposition~\ref{P2}, for any feasible performance requirement $\varepsilon_{\textrm{req}}\in[\varepsilon_\textrm{min},+\infty)$, Problem~\ref{Pb1} reduces from constrained optimization to unconstrained FL social welfare maximization.
\begin{proposition}\label{P2}
Given any non-empty set of client types $\mathcal{I}=\mathcal{I}_{L}\cup\mathcal{I}_{H}$, where $\mathcal{I}_L=\{i\in\mathcal{I}:D_i\leq d\gamma^2/\sigma^2\}$, and $\mathcal{I}_H=\{i\in\mathcal{I}:D_i> d\gamma^2/\sigma^2\}$. In the MoTS framework, maximizing social welfare $W_\textrm{MoTS}(\boldsymbol{K},\boldsymbol{B})$ is equivalent to minimizing FL model generalization error $\varepsilon(\boldsymbol{K})$ under the following conditions:
\begin{itemize}
    \item if $\mathcal{I}_H=\emptyset$: for any non-empty subset $\mathcal{J}\subset\mathcal{I}$, each client type $i\in\mathcal{I}\setminus\mathcal{J}$ with $\Psi_i\geq0$ satisfies
    \begin{align}\label{Ep21}
            \frac{(K+N_i)^2N_iC_i/N}{\partial U\left(\varepsilon_{\mathcal{J}}\right)/\partial K_i}\leq&\Psi_i,
    \end{align}
    where $\Psi_i=\frac{d\gamma^2N_i}{D_i}+N_i(N_i+2K-1)\sigma^2-(2KN_i+1)\varepsilon_{\mathcal{J}}$, $\varepsilon_{\mathcal{J}}$ is the model generalization error trained by all clients indexed by $\mathcal{J}$, and $K=\sum_{j\in\mathcal{J}}K_j$;
    \item if $\mathcal{I}_H\neq\emptyset$: for any non-empty participant coalition $\mathcal{K}$, each client type $i\in\mathcal{I}$ with $\Psi_i\geq0$ satisfies
    \begin{align}\label{Ep22}
            \frac{(K+1)^2C_i/N}{\partial U\left({d\gamma^2}/{D_i}+2K(\sigma^2-\varepsilon_{\mathcal{K}})\right)/\partial K_i}\leq\Psi_i,
    \end{align}
    where $\Psi_i=\frac{d\gamma^2}{D_i}+2K\sigma^2-(2K+1)\varepsilon_{\mathcal{K}}$, $\varepsilon _{\mathcal{K}}$ is the model generalization error trained by $\mathcal{K}$, and $K=\sum_{i\in\mathcal{I}}K_i$.
\end{itemize}
\end{proposition}

We give the proof of Proposition~\ref{P2} in online appendix~\cite{Online_appendix}. Proposition~\ref{P2} provides a theoretical foundation for understanding the relationship between social welfare maximization and model generalization error minimization. From the conditions in~\eqref{Ep21} and~\eqref{Ep22}, the misalignment between these two objectives stems from the trade-off between incremental model utility $\partial U\left(\varepsilon\right)/\partial K_i$ and cost $C_i$, induced by client participation. When heterogeneous clients participate in FL, if the increase in model utility consistently outweighs their participation costs, then social welfare maximization coincides with improving the model, i.e., reducing error $\varepsilon$. This alignment eliminates the need for an explicit error constraint $\varepsilon_{\textrm{req}}\in[\varepsilon_\textrm{min},+\infty)$ in application-aware social welfare maximization, reducing the complexity of Problem~\ref{Pb1}. In the MoTS framework, the larger client population $N$ will amplify the positive externality of model improvements, thereby fostering the alignment between maximizing $W_\textrm{MoTS}(\boldsymbol{K},\boldsymbol{B})$ and minimizing $\varepsilon(\boldsymbol{K})$.

Following the analysis of optimal social states $\{\boldsymbol{K}^*,\boldsymbol{B}^*\}$ for application-aware social welfare maximization, we next address its practical implementation. We aim to design a mechanism that leads self-interested clients, who prioritize optimizing their individual payoff $\pi_n$ and thus may deviate from $\{\boldsymbol{K}^*,\boldsymbol{B}^*\}$, to maximize social welfare $W_\textrm{MoTS}(\boldsymbol{K},\boldsymbol{B})$ while satisfying the generalization error requirement $\varepsilon_{\textrm{req}}$.

\subsection{Social Welfare Maximization with Application-Aware and Network Effects (SWAN) Mechanism Design}
The mechanism design for application-aware social welfare maximization, as formulated in Problem~\ref{Pb1}, involves determining the FL model price $p$ and type-$i$ participation reward $r_i$.

\textbf{Mechanism Design Challenge.} From Lemma~\ref{L1} and Proposition~\ref{P1}, achieving the optimal social states $\{\boldsymbol{K}^*,\boldsymbol{B}^*\}$ for Problem~\ref{Pb1} requires strategic coordination among heterogeneous clients. For example, with low data distribution heterogeneity $\sigma^2\leq d\gamma^2/D_I$, the optimal social states exhibit an ``all-or-none'' participation pattern across all client types. However, each client $n\in\mathcal{N}_i$, irrespective of type $i\in\mathcal{I}$, makes individual decisions to maximize their own payoff $\pi_{n}$ in~\eqref{E6}, without considering social welfare $W_{\textrm{MoTS}}(\boldsymbol{K},\boldsymbol{B})$. This divergence between individual incentives and social optimality, coupled with network effects (i.e., interdependence in client decisions), creates the central challenge for mechanism design.

\textbf{SWAN Mechanism.} To address this challenge, we propose a \underline{S}ocial \underline{W}elfare Maximization with \underline{A}pplication-Aware and \underline{N}etwork Effects (SWAN) mechanism. The SWAN mechanism incorporates non-monotonic network effects (see Theorem~\ref{T2}) into the pricing and reward structures, ensuring that individual payoff maximization leads to desired social states $\{\boldsymbol{K}^*,\boldsymbol{B}^*\}$. Corresponding to Proposition~\ref{P1}, the SWAN mechanism divides the implementation into two scenarios based on distinct data distribution heterogeneity (i.e., non-i.i.d. levels).

(\romannumeral1) For scenarios with low heterogeneity ($\sigma^2\leq d\gamma^2/D_I$), the optimal social states satisfy $K_i^*\in\{0,N_i\}$, for all $i\in\mathcal{I}$. The SWAN mechanism constructs a multilinear function $\theta(\boldsymbol{K})$  through interpolation \cite{stoer1980introduction} to enable transitions between “all-or-none” social states. Specifically,
\begin{align}\label{E8}
    \theta(\boldsymbol{K})=\sum_{\boldsymbol{x}\in\mathcal{X}} L(\boldsymbol{x})\prod_{i=1}^I \frac{(2K_i-N_i)x_i+(N_i-K_i)N_i}{N^2_i},
\end{align}
where $\mathcal{X}=\prod_{i=1}^I\{0,N_i\}$, and $L(\boldsymbol{K})$ is the Lagrangian of Problem~\ref{Pb2}:
\begin{align}\label{eqL}
    L(\boldsymbol{K})=&\sum_{i\in\mathcal{I}} \left[N_i\cdot U\big(\varepsilon(\boldsymbol{K})\big)-K_i\cdot C_i\right]\nonumber\\&+\lambda\left[(\varepsilon(\boldsymbol{K})-\sigma^2)^{-1}-(\varepsilon_{\textrm{req}}-\sigma^2)^{-1}\right],
\end{align}
with $\lambda$ denoting the Lagrange multiplier associated with the error constraint $\varepsilon(\boldsymbol{K}) \leq \varepsilon_{\textrm{req}}$, $\varepsilon_{\textrm{req}}\in[\varepsilon_\textrm{min},+\infty)$.

(\romannumeral2) For scenarios with high heterogeneity ($\sigma^2> d\gamma^2/D_I$), partial participation of clients in FL model training can also be optimal. The SWAN mechanism thus directly employs the Lagrangian $L(\boldsymbol{K})$ in~\eqref{eqL} to support various participation patterns for application-aware social welfare maximization.

To minimize the platform's additional incentive costs while achieving social optimality, we further introduce an incentive ratio $\tau$, which dynamically scales the mechanism's reward and pricing based on the social welfare generated:
\begin{subnumcases}{\label{eq13}}
    \tau=\frac{W_{\textrm{MoTS}}(\boldsymbol{K}^*,\boldsymbol{B}^*)/N}{L(\boldsymbol{K}^*)-L_0},&\textrm{if} $W_{\textrm{MoTS}}(\boldsymbol{K}^*,\boldsymbol{B}^*)>0$, \\
    \tau\rightarrow0^+,&\textrm{if} $W_{\textrm{MoTS}}(\boldsymbol{K}^*,\boldsymbol{B}^*)\leq0$,
\end{subnumcases}
where $L_0=\inf_{\boldsymbol{K}} L(\boldsymbol{K})$ represents the minimum value of the Lagrangian over all feasible participation patterns. The rationale behind this ratio design is that when social welfare is positive, the mechanism can achieve budget balance (zero platform costs) by redistributing the generated welfare among clients. When the optimal social welfare is non-positive, the platform provides minimal incentives to ensure participation.

Building upon the above formulation, we present the complete SWAN mechanism below.
\begin{mechanism}[The SWAN Mechanism]\label{M1}
Consider any set of client types $\mathcal{I}=\mathcal{I}_{L}\cup\mathcal{I}_{H}$, where $\mathcal{I}_L=\{i\in\mathcal{I}:D_i\leq d\gamma^2/\sigma^2\}$, and $\mathcal{I}_H=\{i\in\mathcal{I}:D_i> d\gamma^2/\sigma^2\}$. The SWAN mechanism operates as follows:

(\romannumeral1) For the scenario with low data distribution heterogeneity, i.e., $\sigma^2\leq d\gamma^2/D_I$, which implies that $\mathcal{I}_H=\emptyset$:
\begin{align}
\begin{cases}
    p=U(\varepsilon)-\tau[\theta(\boldsymbol{K})-L_0],\\
    r_{i}= C_i-U(\varepsilon)+\tau[\theta(\boldsymbol{K})-L_0],\quad \forall i\in\mathcal{I}.
\end{cases}
\end{align}

(\romannumeral2) For the scenario with high data distribution heterogeneity, i.e., $\sigma^2>d\gamma^2/D_I$, which implies that $\mathcal{I}_H\neq\emptyset$:
\begin{align}
\begin{cases}
    p=U(\varepsilon)-\tau[L(\boldsymbol{K})-L_0],\\
    r_{i}=C_i-U(\varepsilon)+\tau[L(\boldsymbol{K})-L_0],\quad \forall i\in\mathcal{I}.
\end{cases}
\end{align}
\end{mechanism}

By design, the SWAN mechanism combines heterogeneity-adaptive coordination with welfare-based incentive scaling.

\textbf{Optimality Analysis of SWAN.} With the proposed SWAN mechanism, we next establish its optimality and the dominance of the MoTS framework over existing FL frameworks. We characterize the key properties of SWAN in Theorem~\ref{T3}, and provide a detailed proof in online appendix~\cite{Online_appendix}.
\begin{thm}
    \label{T3}
    The SWAN mechanism achieves the optimal social states $\{\boldsymbol{K}^*,\boldsymbol{B}^*\}$ that maximize $W_{\emph{MoTS}}(\boldsymbol{K},\boldsymbol{B})$ subject to $\varepsilon(\boldsymbol{K}^*) \leq \varepsilon_{\emph{req}}$. Furthermore, when $W_{\emph{MoTS}}(\boldsymbol{K}^*,\boldsymbol{B}^*) > 0$, the mechanism is budget-balanced, incurring zero additional costs, i.e., $\sum_{i\in\mathcal{I}}(B^*_i\cdot p-K^*_i\cdot r_i)=0$.
\end{thm}
From Theorem~\ref{T3}, the SWAN mechanism solves Problem~\ref{Pb1} with two desirable properties: social optimality and budget balance (i.e., zero additional costs). The designed pricing $p$ and reward $r_i$ internalize network effects, aligning each client's optimal decision $s_n^*$ with maximizing $W_{\textrm{MoTS}}(\boldsymbol{K},\boldsymbol{B})$ under error requirement $\varepsilon_{\textrm{req}}$. At the optimal social states $\{\boldsymbol{K}^*,\boldsymbol{B}^*\}$, the SWAN mechanism minimizes the platform's additional incentive costs, and achieves budget balance whenever optimal social welfare $W_{\textrm{MoTS}}(\boldsymbol{K}^*,\boldsymbol{B}^*)>0$.

Following Theorem~\ref{T3}, we compare the optimal social welfare $W_{\textrm{MoTS}}(\boldsymbol{K}^*,\boldsymbol{B}^*)$ under the MoTS framework against that under the existing FL framework. The existing FL framework (e.g., \cite{mcmahan2017communication,Tan2022,Abdul2021}) restricts model access to training participants, thereby yielding social welfare:
\begin{equation}\label{eqWFL}
    W_{\textrm{FL}}(\boldsymbol{K})=\sum_{i\in\mathcal{I}}K_i\cdot (U(\varepsilon)- C_i).
\end{equation}

In contrast, the MoTS framework enables non-participating clients to purchase the trained global FL model, resulting in $W_{\textrm{MoTS}}(\boldsymbol{K},\boldsymbol{B})$ in~\eqref{E5}. Consequently, the MoTS framework achieves no less social welfare than the existing FL framework at optimality, as stated in Proposition~\ref{P3}.

\begin{proposition}\label{P3}
For any error requirement $\varepsilon_\emph{req}\in[\varepsilon_\emph{min},+\infty)$, the MoTS framework achieves $W_\emph{MoTS}(\boldsymbol{K}^*,\boldsymbol{B}^*)\geq W_\emph{FL}(\boldsymbol{K}^*)$, where $W_{\emph{FL}}(\boldsymbol{K}^*)$ indicates the maximum achievable social welfare under the existing FL framework with identical model performance $\varepsilon(\boldsymbol{K}^*)$.
\end{proposition}

To summarize, we have analyzed the network effects of client participation and designed the SWAN mechanism for application-aware social welfare maximization. Next, we empirically evaluate our theoretical results through extensive experiments on an FL hardware prototype.

\section{Experimental Evaluation}
\label{S4}
This section presents experiments to validate our theoretical analysis and demonstrate the practical effectiveness of our proposed mechanism. We first introduce the experimental setup, then empirically characterize the network effects of client participation, and finally evaluate the performance of our SWAN mechanism and the MoTS framework.

\subsection{Experimental Setup}
\subsubsection{Hardware Prototype of FL} 
\begin{figure}[!htbp]
    \centering
    \includegraphics[width=0.35\textwidth]{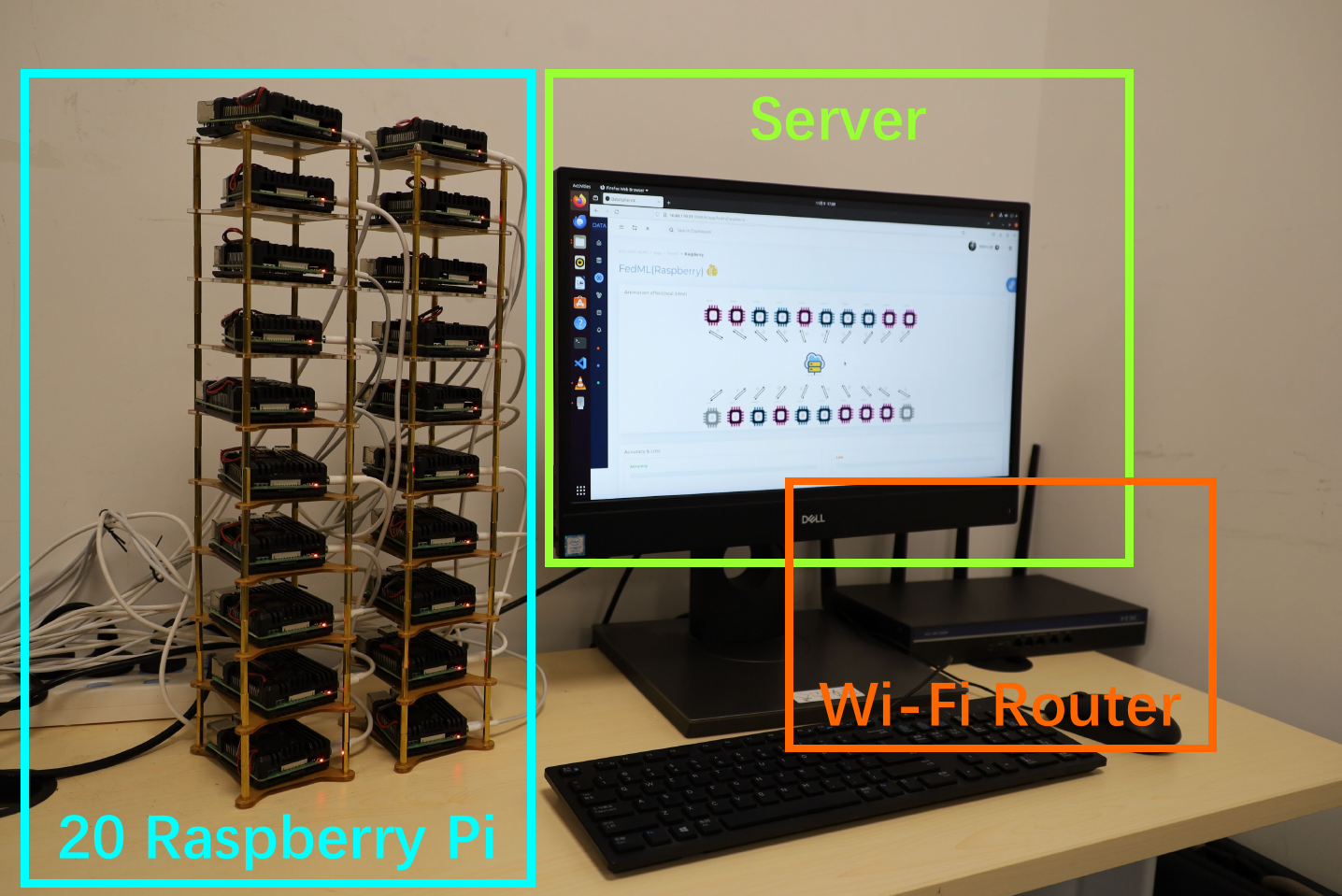}
    \caption{Hardware Prototype of FL System with 20 Raspberry Pi Clients.}
    \label{HD}
    \vspace{-4pt}
\end{figure}

To better reflect realistic deployment scenarios, we implement a hardware-based FL testbed, as illustrated in Figure~\ref{HD}.  The testbed comprises $20$ Raspberry Pi devices as clients and a desktop server for coordination, connected via Wi-Fi with TCP-based communication. This hardware setup captures real-world deployment constraints, including network latency, computational limitations, and communication overhead.
\subsubsection{Heterogeneous Clients and Learning Implementation} 
We configure $N=20$ heterogeneous clients across $I=3$ types: $10$ type-$1$ clients, five type-$2$ clients, and five type-$3$ clients. For concreteness, we set the FL model utility function\footnote{The key insights remain consistent across other functional forms of model utility and participation cost.} as $U(\varepsilon)=40\cdot\varepsilon^{-16}$, and participation cost proportional to training data size~\cite{karimireddy2022mechanisms}. While our experiments focus on this small-scale setup, the insights generalize to larger and more diverse deployments.

We conduct experiments on three widely adopted datasets\footnote{We also conduct experiments on the CIFAR-$100$~\cite{krizhevsky2009learning} dataset, which yield similar results, as shown in the online appendix~\cite{Online_appendix}.}: MNIST~\cite{Lecun}, SVHN~\cite{37648}, and CIFAR-$10$~\cite{krizhevsky2009learning}, with different model architectures and data distributions. Table~\ref{CF10} summarizes the experimental configurations. For all experiments, we evaluate the aggregated model at each communication round and use the test loss as the generalization error $\varepsilon$ of the trained global FL model. To ensure reliability, all results are averaged over $30$ independent runs. Below, we detail the implementation settings for each dataset.
\begin{table}[!htbp]\vspace{-4pt}
\caption{Experimental Configuration Across Datasets.}
\label{CF10}
\centering
\begin{tabular}{llccc}
\toprule
\multicolumn{2}{c}{\textbf{Configuration}} & \textbf{MNIST} & \textbf{SVHN} & \textbf{CIFAR-10} \\
\midrule
\multicolumn{2}{l}{Test Size} & 10,000 & 26,000 & 10,000 \\
\midrule
\multirow{3}{*}{Training Size} 
    & $D_1$ & 50 & 1,400 & 1,000 \\
    & $D_2$ & 120 & 3,200 & 2,250 \\
    & $D_3$ & 300 & 8,000 & 5,750 \\
\midrule
\multicolumn{2}{l}{Data Distribution} & i.i.d. & i.i.d. & non-i.i.d. \\
\midrule
\multicolumn{2}{l}{FL Model} & \makecell{Logistic\\(Convex)} & \makecell{CNN\\(Non-convex)} & \makecell{CNN\\(Non-convex)} \\
\bottomrule
\end{tabular}
\end{table}

\textbf{Setup with MNIST Dataset:} In the empirical experiments on the MNIST dataset, we follow the same setup as in our theoretical analysis, and study extreme scenarios without data distribution heterogeneity (i.e., i.i.d. data distribution, where client variance $\sigma^2=0$). We adopt a convex multinomial logistic regression model, with initialized parameters $\boldsymbol{w}_0=\boldsymbol{0}$ and stochastic gradient descent (SGD) batch size $b=32$. Similar to \cite{Li2020On}, we use an initial learning rate $\eta_0=0.001$ with decay rate ${\eta_0}/({1+r})$, where $r$ is the communication round index, and the local iteration number $E=20$. The total number of communication rounds is $500$.

\textbf{Setup with SVHN and CIFAR-10 Datasets:} In the empirical experiments on the SVHN and CIFAR-$10$ datasets, we generalize from our theoretical setting to explore more complex and realistic non-convex models. We simulate an idealized FL scenario with i.i.d. data distribution on the SVHN dataset, while adopting a heterogeneous (non-i.i.d.) data distribution across clients on the CIFAR-$10$ dataset. Specifically, in the non-i.i.d. scenario, we assign one class of data samples to type-$1$ clients, five classes to type-$2$ clients, and all $10$ classes to type-$3$ clients. For both datasets, we employ a convolutional neural network (CNN) model, which consists of two convolutional layers (ReLU activation function), three fully connected layers, and a max pooling layer applied to each convolutional layer. We initialize the CNN model with parameters $\boldsymbol{w}_0=\boldsymbol{0}$ and set the SGD batch size to $b=64$. The initial learning rate is $\eta_0=0.1$ with the decay rate ${\eta_0}/({1+r})$, and the local iteration number $E=30$. The total number of communication rounds is $600$.

\subsection{Validation of Network Effects}
\begin{figure}[ht]
    \centering
    \subfloat[Convex FL Model (MNIST).]{\label{N_mnist}\includegraphics[width=0.5\linewidth]{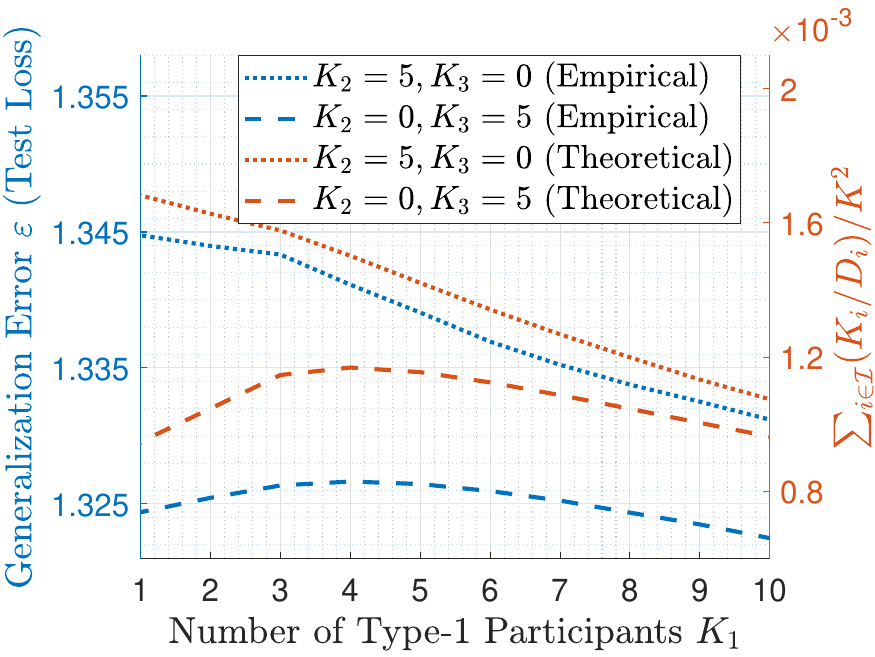}}\hfill
    \subfloat[Non-convex FL Model (SVHN).]{\label{N_svhn}\includegraphics[width=0.5\linewidth]{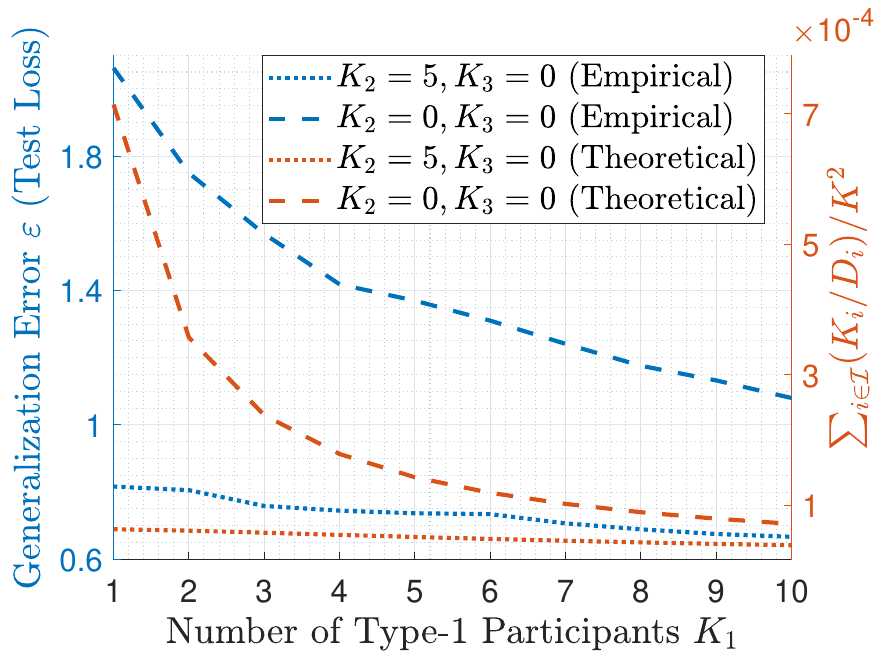}}
    \caption{Network Effects of Client Participation: Empirical Results (Blue) versus Theoretical Predictions (Red).}
    \label{N_MS}
\end{figure}

Figure~\ref{N_MS} demonstrates network effects predicted by Theorem~\ref{T2}, comparing empirical results (blue curves, left y-axis) with theoretical predictions (red curves, right y-axis) from the generalization error $\varepsilon$ derived in~\eqref{E7}.

We first examine the linear model setup in Figure~\ref{N_MS}\subref{N_mnist}, which directly corresponds to our theoretical analysis. The empirical results reveal that network effects vary with participant composition. When existing participants are of type-$2$ ($K_2=5$, $K_3=0$, dotted curves), the participation of type-$1$ clients consistently reduces the generalization error, exhibiting positive network effects. In contrast, when type-$3$ participants dominate ($K_2=0$, $K_3=5$, dashed curves), increasing $K_1$ initially raises the error before eventually reducing it, indicating a transition from negative to positive network effects. This non-monotonic pattern aligns with the region partition of network effects in Figure~\ref{Region} and Theorem~\ref{T2}.
\begin{observation}[Non-Monotonic Network Effects]
Consistent with Theorem~\ref{T2}, FL network effects are non-monotonic, transitioning between positive and negative based on the relative data sizes and heterogeneity levels across participants.
\end{observation}

The strong alignment between empirical (blue) and theoretical (red) curves further validates our analytical model. For example, in Figure~\ref{N_MS}\subref{N_mnist}, both show that the generalization error peaks at $K_1=4$ when $K_2=0$, $K_3=5$. Figure~\ref{N_MS}\subref{N_svhn} extends validation to the non-convex FL model, which differs from our theoretical setting. Nevertheless, the empirical results remain aligned with the predicted trends of $\varepsilon$ in~\eqref{E7}.
\begin{observation}[Theoretical Validity and Generalizability]
The comparison between empirical and theoretical curves illustrates strong alignment, which validates our analytical model and suggests its generalizability to non-convex settings.
\end{observation}

\subsection{Evaluation of SWAN Mechanism and MoTS Framework}
We evaluate the performance of the proposed SWAN mechanism and MoTS framework in Figures~\ref{F_CM1}--\ref{F_CF2}, comparing against the following benchmarks:
\begin{itemize}
    \item \textit{Dynamic Mechanism}~\cite{hu2023}: Considers clients' evolving expectations of FL model performance through network effects games. However, incomplete information and client heterogeneity could cause it to deviate from application-aware social welfare maximization.
    \item \textit{Modified FL Mechanism}~\cite{Yan22}: A modified version of the existing FL incentive mechanism adapted to the MoTS framework. It sets the model price equal to the FL model utility but provides fixed participation rewards, failing to address the network effects of client participation.
    \item \textit{Theoretical Optimum of FL Framework}: Represents the maximum achievable performance under existing FL frameworks~\cite{mcmahan2017communication,Tan2022,Abdul2021}. This theoretical upper bound provides a critical baseline for evaluating the effectiveness of our MoTS framework.
\end{itemize}
\subsubsection{The Performance of SWAN Mechanism} 
\begin{figure}[!htbp]
    \centering
    \subfloat[Varying Participation Cost (No Error Constraint, i.e., $\varepsilon_{\textrm{req}}=+\infty$).]{\label{C_mnist_R=inf}\includegraphics[width=0.49\linewidth]{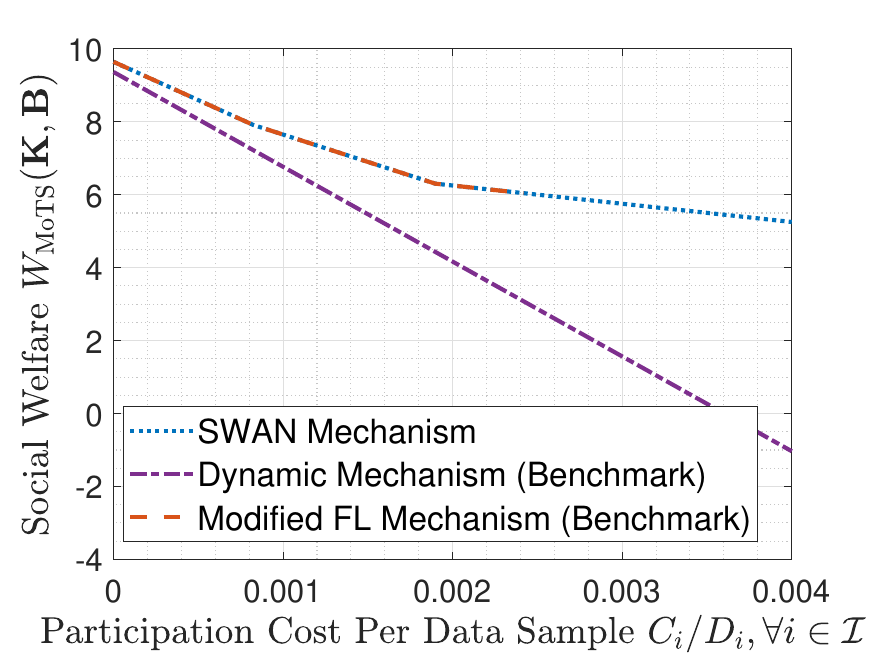}}\hfill
    \subfloat[Varying Error Requirement (Participation Cost $C_i/D_i=0.002$).]{\label{R_mnist_C=0.002}\includegraphics[width=0.49\linewidth]{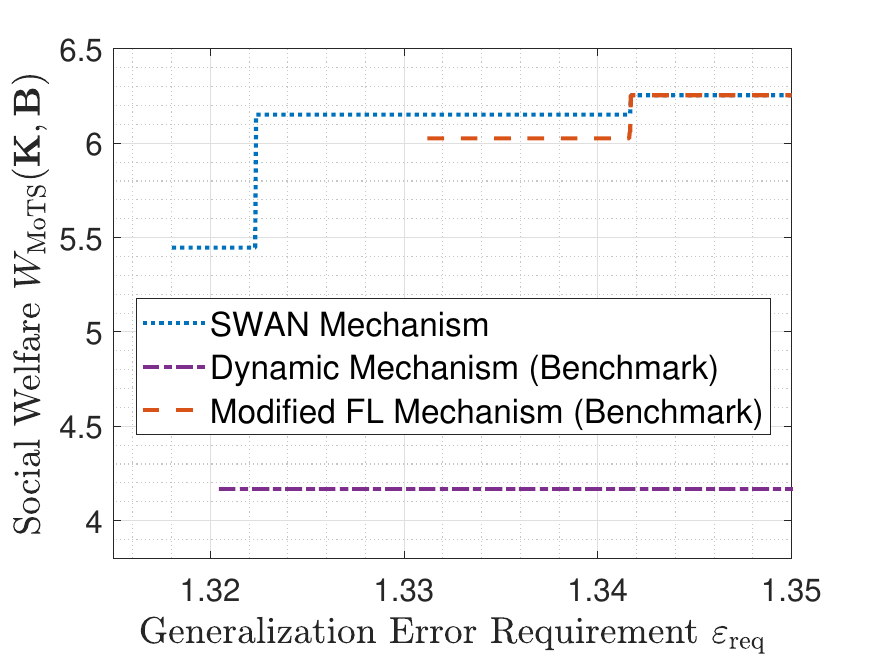}}
    \caption{SWAN versus benchmark mechanisms: social welfare $W_\textrm{MoTS}(\boldsymbol{K},\boldsymbol{B})$ under the MoTS framework (convex model, i.i.d. data, MNIST).}
    \label{F_CM1}
\end{figure}

\begin{figure}[!htbp]
    \centering
    \subfloat[Varying Participation Cost (No Error Constraint, i.e., $\varepsilon_{\textrm{req}}=+\infty$).]{\label{C_cifar_R=inf}\includegraphics[width=0.49\linewidth]{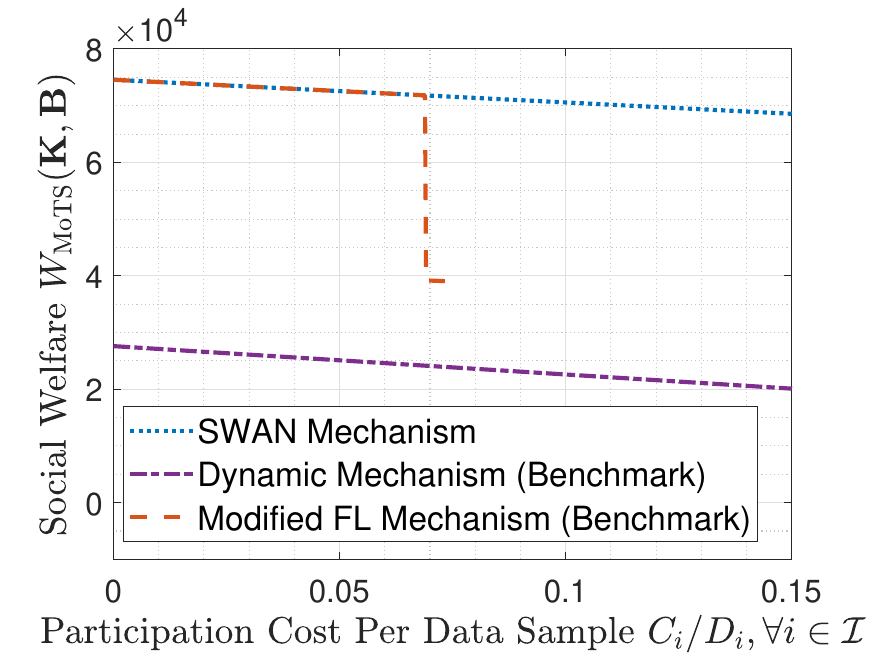}}\hfill
    \subfloat[Varying Error Requirement (Participation Cost $C_i/D_i=0.07$).]{\label{R_cifar_C=0.07}\includegraphics[width=0.49\linewidth]{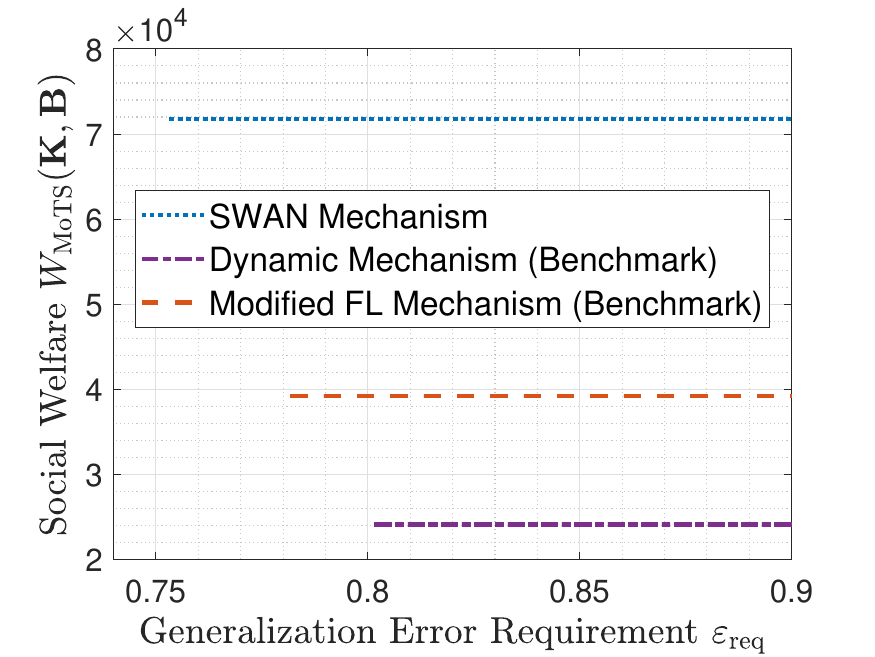}}
    \caption{SWAN versus benchmark mechanisms: social welfare $W_\textrm{MoTS}(\boldsymbol{K},\boldsymbol{B})$ under the MoTS framework (non-convex model, non-i.i.d. data, CIFAR-10).}
    \label{F_CM2}
\end{figure}

In Figure~\ref{F_CM1} and Figure~\ref{F_CM2}, we compare SWAN with the dynamic mechanism and modified FL mechanism, all operating under the MoTS framework, to evaluate their social welfare across varying participation costs and error requirements $\varepsilon_{\textrm{req}}\in[\varepsilon_\textrm{min},+\infty)$.

Figure~\ref{F_CM1} presents experimental results for the convex FL model with i.i.d. client data. We examine how per-sample participation cost $C_i/D_i$ and application-specific error requirement $\varepsilon_{\textrm{req}}$ affect the social welfare $W_{\textrm{MoTS}}(\boldsymbol{K},\boldsymbol{B})$ under different mechanisms. From Figure \ref{F_CM1}\subref{C_mnist_R=inf}, when $\varepsilon_{\textrm{req}}=+\infty$ (i.e., without generalization error constraint), $W_{\textrm{{MoTS}}}(\boldsymbol{K},\boldsymbol{B})$ declines with increasing participation cost across all mechanisms. Nevertheless, SWAN exhibits superior performance compared to both benchmarks. Specifically, the benchmarks achieve competitive social welfare only at low participation costs. As the cost increases, the dynamic mechanism suffers rapid degradation while the modified FL mechanism fails entirely to incentivize clients, due to client heterogeneity and negative network effects of client participation. In contrast, SWAN incentivizes heterogeneous clients to overcome temporary negative network effects and short-term payoff losses, achieving higher social welfare across all cost levels.

Figure~\ref{F_CM1}\subref{R_mnist_C=0.002} further shows that SWAN accommodates a wider range of generalization error requirements $\varepsilon_{\textrm{req}}$ than the benchmarks. This adaptability enables SWAN to satisfy stringent error demands imposed by diverse applications while maximizing social welfare, whereas the benchmarks operate only under relaxed error constraints. For instance, the modified FL mechanism becomes effective only when $\varepsilon_{\textrm{req}}\geq 1.332$.
\begin{observation}[Cost Robustness and Application Adaptability]
SWAN demonstrates superior robustness to participation costs and accommodates broader generalization error requirements than benchmarks, which either degrade rapidly under elevated costs or fail under strict error constraints.
\end{observation}

Figure~\ref{F_CM2} extends the evaluation to a more realistic setting with a non-convex FL model and non-i.i.d. client data. Consistent with Figure~\ref{F_CM1}, SWAN maintains its performance advantage over both benchmarks. Furthermore, the increased client statistical heterogeneity makes SWAN's advantage more pronounced: compared to the i.i.d. scenario, SWAN outperforms the benchmarks by a larger margin in $W_{\textrm{{MoTS}}}(\boldsymbol{K},\boldsymbol{B})$ and supports a wider feasible range of $\varepsilon_{\textrm{req}}$.
\begin{observation}[Effectiveness in Practical FL Settings]
In non-convex settings with non-i.i.d. data, SWAN's advantages over the benchmarks persist and are more pronounced.
\end{observation}
\subsubsection{The Performance of MoTS Framework} 
\begin{figure}[!htbp]
    \centering
    \subfloat[Varying Participation Cost (Error Requirement $\varepsilon_{\textrm{req}}=1.35$).]{\label{C_mnist_SvI_R=1.35}\includegraphics[width=0.49\linewidth]{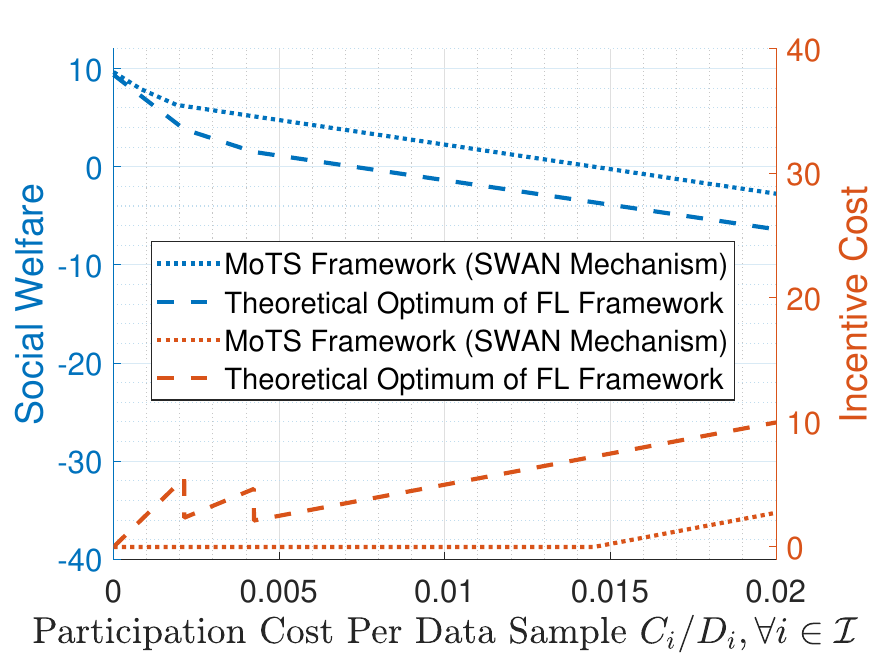}}\hfill
    \subfloat[Varying Error Requirement (Participation Cost $C_i/D_i=0.01$).]{\label{R_mnist_SvI_C=0.01}\includegraphics[width=0.48\linewidth]{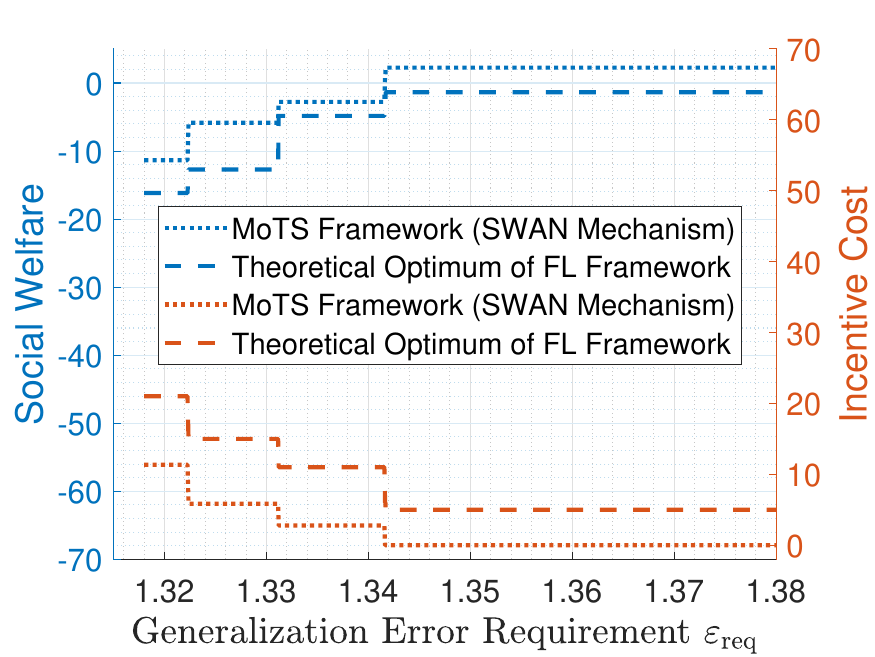}}
    \caption{MoTS versus FL framework: social welfare and incentive cost comparison (convex model, i.i.d. client data, MNIST).}
    \label{F_CF1}
\end{figure}

\begin{figure}[!htbp]
    \centering
    \subfloat[Varying Participation Cost (Error Requirement $\varepsilon_{\textrm{req}}=0.8$).]{\label{C_cifar_SvI_R=1.8}\includegraphics[width=0.49\linewidth]{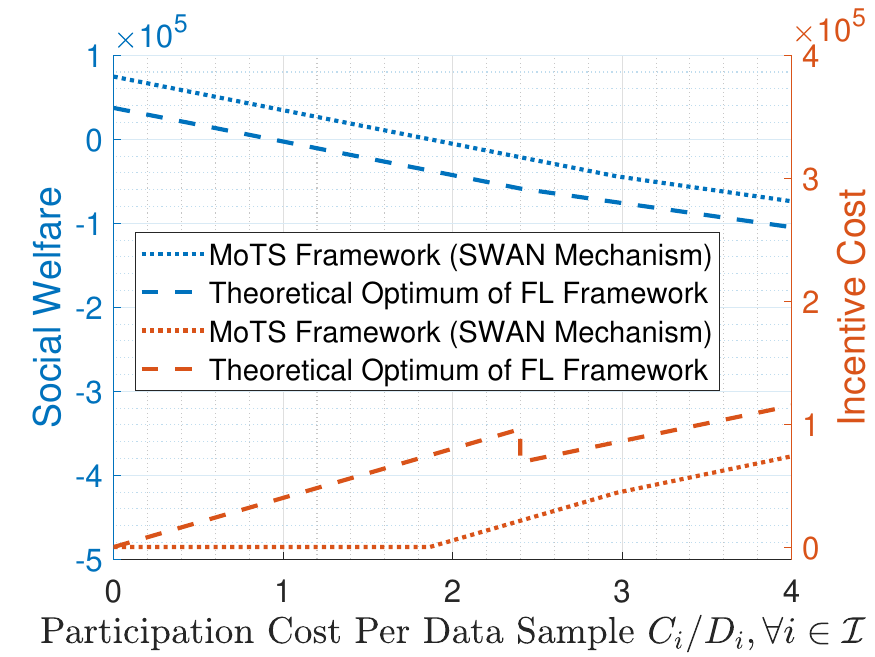}}\hfill
    \subfloat[Varying Error Requirement (Participation Cost $C_i/D_i=1.8$).]{\label{R_CIFAR_SvI_C=0.8}\includegraphics[width=0.49\linewidth]{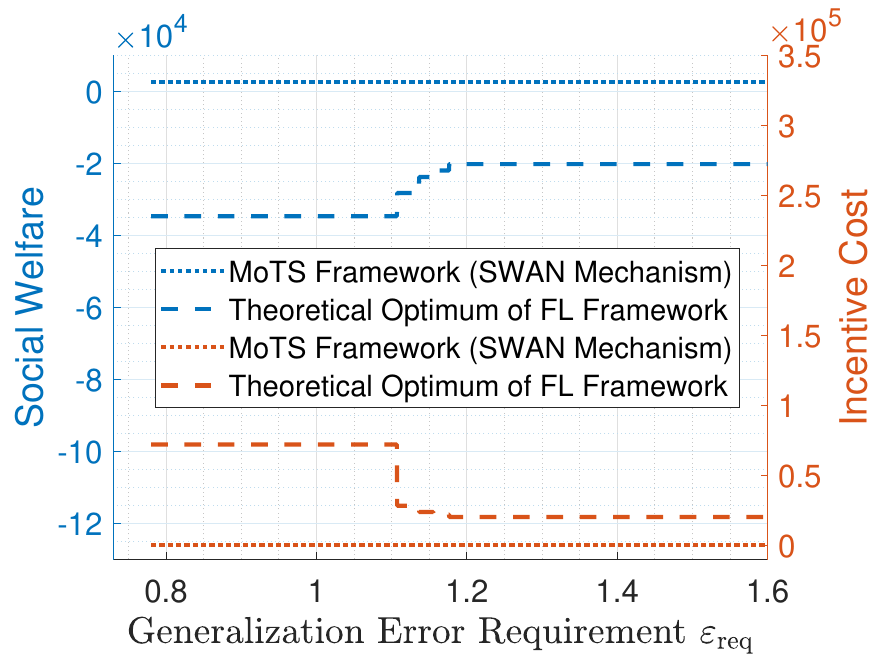}}
    \caption{MoTS versus FL framework: social welfare and incentive cost comparison (non-convex model, non-i.i.d. client data, CIFAR-10).}
    \label{F_CF2}
\end{figure}

In Figure~\ref{F_CF1} and Figure~\ref{F_CF2}, we compare the MoTS framework with the theoretical optimum of existing FL frameworks~\cite{mcmahan2017communication,Tan2022,Abdul2021}, evaluating both social welfare (blue, left y-axis) and platform incentive costs (red, right y-axis).

Figure~\ref{F_CF1} shows the performance of our MoTS framework against the existing FL framework under the convex model with i.i.d. client data. As the participation cost $C_i/D_i$ increases or the error requirement $\varepsilon_{\textrm{req}}$ becomes more stringent, social welfare decreases and incentive costs rise for both frameworks. However, by introducing a purchasing option and leveraging network effects through SWAN, the MoTS framework significantly outperforms the FL framework. Specifically, in Figure~\ref{F_CF1}\subref{C_mnist_SvI_R=1.35}, under varying participation costs, MoTS achieves $352.42\%$ higher social welfare while reducing incentive costs by $93.07\%$. In Figure~\ref{F_CF1}\subref{R_mnist_SvI_C=0.01}, under varying error requirements, MoTS achieves $100.48\%$ higher social welfare with $79.38\%$ lower platform incentive costs.

Figure~\ref{F_CF2} examines the more challenging setting of a non-convex FL model with non-i.i.d. client data. Consistent with Figure~\ref{F_CF1}, the MoTS framework continues to outperform the existing FL framework. As shown in Figure \ref{F_CF2}\subref{C_cifar_SvI_R=1.8}, under varying participation costs $C_i/D_i$, MoTS achieves $89.94\%$ higher welfare and $67.51\%$ lower costs. For varying error requirements $\varepsilon_{\textrm{req}}$ in Figure~\ref{F_CF2}\subref{R_CIFAR_SvI_C=0.8}, MoTS achieves $109.72\%$ higher welfare while eliminating incentive costs entirely. Notably, MoTS incurs zero incentive cost once social welfare becomes positive, which validates Theorem~\ref{T3}. Moreover, in Figure~\ref{F_CF2}\subref{R_CIFAR_SvI_C=0.8}, MoTS exhibits constant social welfare and incentive cost across different $\varepsilon_{\textrm{req}}$, confirming the conditions in Proposition~\ref{P2}, where social welfare maximization aligns with generalization error minimization.

\begin{observation}[Superior Efficiency of MoTS Framework]
Compared to the theoretical optimum of existing FL frameworks, MoTS achieves significantly higher social welfare (up to $352.42\%$) while reducing platform costs (up to $100\%$).
\end{observation}
\begin{observation}[Budget Balance and Welfare-Performance Alignment]
MoTS achieves zero platform costs when social welfare is positive, validating Theorem~\ref{T3}. Under conditions in Proposition~\ref{P2}, where social welfare maximization aligns with generalization error minimization, MoTS maintains the maximum social welfare across varying $\varepsilon_{\textrm{req}}\in[\varepsilon_\textrm{min},+\infty)$.
\end{observation}

\section{Conclusion}
\label{S5}
This work provides the first study on maximizing social welfare while satisfying application-specific performance (i.e., generalization error) requirements in FL with network effects. Our analysis reveals that the network effects of client participation are not monotonic, which depends on client heterogeneity. To address this, we propose the MoTS framework for FL, enabling application-aware social welfare optimization. Within the MoTS framework, we further design a SWAN mechanism, which leverages network effects to incentivize heterogeneous clients through payments from model customers, minimizing extra incentive costs. The SWAN mechanism flexibly adapts to diverse performance requirements across FL applications while maximizing social welfare. Once the achieved social welfare is positive, the SWAN mechanism operates without necessitating any additional incentive costs.

In future works, we will extend our incentive mechanism to account for clients' strategic data contribution decisions, e.g., limited participation using only part of their private data. Furthermore, we will refine our theoretical model by incorporating clients' utility and computational heterogeneity. Such advancements, while significantly complicating the equilibrium analysis and mechanism design, are expected to enhance the practicality of our research.


\bibliographystyle{IEEEtran}
\bibliography{ref}

@inproceedings{10.1145/3641512.3686394,
author = {Li, Xiang and Luo, Yuan and Luo, Bing and Huang, Jianwei},
title = {Social Welfare Maximization for Federated Learning with Network Effects},
year = {2024},
booktitle = {International Symposium on Theory, Algorithmic Foundations, and Protocol Design for Mobile Networks and Mobile Computing},
pages = {131-140},
numpages = {10}
}

@article{10.1016/j.neucom.2024.128019,
author = {Liu, Bingyan and Lv, Nuoyan and Guo, Yuanchun and Li, Yawen},
title = {Recent advances on federated learning: A systematic survey},
year = {2024},
volume = {597},
number = {C},
pages={1-16},
journal = {Neurocomputing}
}

@ARTICLE{Zhan22,
  author={Zhan, Yufeng and Zhang, Jie and Hong, Zicong and Wu, Leijie and Li, Peng and Guo, Song},
  journal={IEEE Transactions on Emerging Topics in Computing}, 
  title={A Survey of Incentive Mechanism Design for Federated Learning}, 
  year={2022},
  volume={10},
  number={2},
  pages={1035-1044}}

@ARTICLE{Tu22,
  author={Tu, Xuezhen and Zhu, Kun and Luong, Nguyen Cong and Niyato, Dusit and Zhang, Yang and Li, Juan},
  journal={IEEE Transactions on Cognitive Communications and Networking}, 
  title={Incentive Mechanisms for Federated Learning: From Economic and Game Theoretic Perspective}, 
  year={2022},
  volume={8},
  number={3},
  pages={1566-1593}}

@ARTICLE{Khan20,
  author={Khan, Latif U. and Pandey, Shashi Raj and Tran, Nguyen H. and Saad, Walid and Han, Zhu and Nguyen, Minh N. H. and Hong, Choong Seon},
  journal={IEEE Communications Magazine}, 
  title={Federated Learning for Edge Networks: Resource Optimization and Incentive Mechanism}, 
  year={2020},
  volume={58},
  number={10},
  pages={88-93}}

@ARTICLE{Nguyen2021-zp,
  author={Nguyen, Dinh C. and Ding, Ming and Pathirana, Pubudu N. and Seneviratne, Aruna and Li, Jun and Vincent Poor, H.},
  journal={IEEE Communications Surveys \& Tutorials}, 
  title={Federated Learning for Internet of Things: A Comprehensive Survey}, 
  year={2021},
  volume={23},
  number={3},
  pages={1622-1658}
  }

@ARTICLE{Guo2021-bf,
  author={Guo, Fengxian and Yu, F. Richard and Zhang, Heli and Li, Xi and Ji, Hong and Leung, Victor C. M.},
  journal={IEEE Internet of Things Journal}, 
  title={Enabling Massive IoT Toward 6G: A Comprehensive Survey}, 
  year={2021},
  volume={8},
  number={15},
  pages={11891-11915}
}

@ARTICLE{10403801,
  author={Sun, Peng and Liao, Guocheng and Chen, Xu and Huang, Jianwei},
  journal={IEEE/ACM Transactions on Networking}, 
  title={A Socially Optimal Data Marketplace With Differentially Private Federated Learning}, 
  year={2024},
  volume={32},
  number={3},
  pages={2221-2236}}

@ARTICLE{9795863,
  author={Pang, Jinlong and Yu, Jieling and Zhou, Ruiting and Lui, John C.S.},
  journal={IEEE Transactions on Mobile Computing}, 
  title={An Incentive Auction for Heterogeneous Client Selection in Federated Learning}, 
  year={2023},
  volume={22},
  number={10},
  pages={5733-5750}}

@ARTICLE{11164975,
  author={Zheng, Zhaohua and Hong, Yiming and Huang, Yue and Qiu, Tie and Xie, Xin and Li, Keqiu},
  journal={IEEE Internet of Things Journal}, 
  title={Uncertainty-Aware Multidimensional Auctions for Social Welfare Optimization in Federated Learning}, 
  year={2025},
  volume={12},
  number={23},
  pages={50454-50465}}

@ARTICLE{Rong21,
  author       = {Rongfei Zeng and
                  Chao Zeng and
                  Xingwei Wang and
                  Bo Li and
                  Xiaowen Chu},
  title        = {A Comprehensive Survey of Incentive Mechanism for Federated Learning},
  journal={arXiv preprint arXiv:2106.15406},
  primaryClass={cs.CL},
  year         = {2021}
}

@ARTICLE{10262054,
  author={Javeed, Danish and Saeed, Muhammad Shahid and Kumar, Prabhat and Jolfaei, Alireza and Islam, Shareeful and Islam, A. K. M. Najmul},
  journal={IEEE Transactions on Consumer Electronics}, 
  title={Federated Learning-Based Personalized Recommendation Systems: An Overview on Security and Privacy Challenges}, 
  year={2024},
  volume={70},
  number={1},
  pages={2618-2627}
}

@ARTICLE{10288131,
  author={Chaddad, Ahmad and Wu, Yihang and Desrosiers, Christian},
  journal={IEEE Internet of Things Journal}, 
  title={Federated Learning for Healthcare Applications}, 
  year={2024},
  volume={11},
  number={5},
  pages={7339-7358}}

@article{10.1145/3501296,
author = {Nguyen, Dinh C. and Pham, Quoc-Viet and Pathirana, Pubudu N. and Ding, Ming and Seneviratne, Aruna and Lin, Zihuai and Dobre, Octavia and Hwang, Won-Joo},
title = {Federated Learning for Smart Healthcare: A Survey},
year = {2022},
volume = {55},
number = {3},
journal = {ACM Computing Surveys},
pages={1-37}
}

@ARTICLE{zhao2023truthful,
      title={Truthful Incentive Mechanism for Federated Learning with Crowdsourced Data Labeling}, 
      author={Yuxi Zhao and Xiaowen Gong and Shiwen Mao},
      year={2023},
      journal={arXiv preprint arXiv:2302.00106},
      primaryClass={cs.LG}
}

@ARTICLE{yang2021achievinglinearspeeduppartial,
      title={Achieving Linear Speedup with Partial Worker Participation in Non-IID Federated Learning}, 
      author={Haibo Yang and Minghong Fang and Jia Liu},
      year={2021},
      journal={arXiv preprint arXiv:2101.11203},
      primaryClass={cs.LG},
}

@INPROCEEDINGS{mcmahan2017communication,
  title={Communication-efficient learning of deep networks from decentralized data},
  booktitle={International Conference on Artificial Intelligence and Statistics},
  author = 	 {McMahan, Brendan and Moore, Eider and Ramage, Daniel and Hampson, Seth and Arcas, Blaise Aguera y},
  pages = 	 {1273--1282},
  year={2017}
}

@ARTICLE{Tan2022,
  author={Tan, Alysa Ziying and Yu, Han and Cui, Lizhen and Yang, Qiang},
  journal={IEEE Transactions on Neural Networks and Learning Systems}, 
  title={Towards Personalized Federated Learning}, 
  year={2023},
  volume={34},
  number={12},
  pages={9587-9603}
}

@ARTICLE{Abdul2021,
  author={Abdulrahman, Sawsan and Tout, Hanine and Ould-Slimane, Hakima and Mourad, Azzam and Talhi, Chamseddine and Guizani, Mohsen},
  journal={IEEE Internet of Things Journal}, 
  title={A Survey on Federated Learning: The Journey From Centralized to Distributed On-Site Learning and Beyond}, 
  year={2021},
  volume={8},
  number={7},
  pages={5476-5497}
}

@book{stoer1980introduction,
  title={Introduction to numerical analysis},
  author={Stoer, Josef and Bulirsch, Roland and Bartels, R and Gautschi, Walter and Witzgall, Christoph},
  volume={1993},
  year={1980},
  publisher={Springer}
}

@ARTICLE{Thi21,
  author={Thi Le, Tra Huong and Tran, Nguyen H. and Tun, Yan Kyaw and Nguyen, Minh N. H. and Pandey, Shashi Raj and Han, Zhu and Hong, Choong Seon},
  journal={IEEE Transactions on Wireless Communications}, 
  title={An Incentive Mechanism for Federated Learning in Wireless Cellular Networks: An Auction Approach}, 
  year={2021},
  volume={20},
  number={8},
  pages={4874-4887}}

@ARTICLE{Lin22,
  author={Lin, Xi and Wu, Jun and Bashir, Ali Kashif and Li, Jianhua and Yang, Wu and Piran, Md. Jalil},
  journal={IEEE Internet of Things Journal}, 
  title={Blockchain-Based Incentive Energy-Knowledge Trading in IoT: Joint Power Transfer and AI Design}, 
  year={2022},
  volume={9},
  number={16},
  pages={14685-14698}}

@ARTICLE{Lee20,
  author={Lee, Joohyung and Kim, DaeJin and Niyato, Dusit},
  journal={IEEE Internet of Things Journal}, 
  title={Market Analysis of Distributed Learning Resource Management for Internet of Things: A Game-Theoretic Approach}, 
  year={2020},
  volume={7},
  number={9},
  pages={8430-8439}}

@ARTICLE{Jiao21,
  author={Jiao, Yutao and Wang, Ping and Niyato, Dusit and Lin, Bin and Kim, Dong In},
  journal={IEEE Transactions on Mobile Computing}, 
  title={Toward an Automated Auction Framework for Wireless Federated Learning Services Market}, 
  year={2021},
  volume={20},
  number={10},
  pages={3034-3048}
  }

@ARTICLE{10292582,
  author={Rodio, Angelo and Faticanti, Francescomaria and Marfoq, Othmane and Neglia, Giovanni and Leonardi, Emilio},
  journal={IEEE/ACM Transactions on Networking}, 
  title={Federated Learning Under Heterogeneous and Correlated Client Availability}, 
  year={2024},
  volume={32},
  number={2},
  pages={1451-1460}}

@ARTICLE{Saputra23,
  author={Saputra, Yuris Mulya and Hoang, Dinh Thai and Nguyen, Diep N. and Tran, Le-Nam and Gong, Shimin and Dutkiewicz, Eryk},
  journal={IEEE Transactions on Mobile Computing}, 
  title={Dynamic Federated Learning-Based Economic Framework for Internet-of-Vehicles}, 
  year={2023},
  volume={22},
  number={4},
  pages={2100-2115}}

@ARTICLE{Zhan20,
  author={Zhan, Yufeng and Li, Peng and Qu, Zhihao and Zeng, Deze and Guo, Song},
  journal={IEEE Internet of Things Journal}, 
  title={A Learning-Based Incentive Mechanism for Federated Learning}, 
  year={2020},
  volume={7},
  number={7},
  pages={6360-6368}
}

@INPROCEEDINGS {Luo23,
author = {B. Luo and Y. Feng and S. Wang and J. Huang and L. Tassiulas},
booktitle = {International Conference on Distributed Computing Systems},
title = {Incentive Mechanism Design for Unbiased Federated Learning with Randomized Client Participation},
year = {2023},
volume = {},
pages = {1-11}
}

@ARTICLE{9797864,
  author={Shi, Zhuan and Zhang, Lan and Yao, Zhenyu and Lyu, Lingjuan and Chen, Cen and Wang, Li and Wang, Junhao and Li, Xiang-Yang},
  journal={IEEE Transactions on Big Data}, 
  title={FedFAIM: A Model Performance-Based Fair Incentive Mechanism for Federated Learning}, 
  year={2024},
  volume={10},
  number={6},
  pages={1038-1050}}

@ARTICLE{9843871,
  author={Zeng, Rongfei and Zeng, Chao and Wang, Xingwei and Li, Bo and Chu, Xiaowen},
  journal={IEEE Network}, 
  title={Incentive Mechanisms in Federated Learning and A Game-Theoretical Approach}, 
  year={2022},
  volume={36},
  number={6},
  pages={229-235}}

@BOOK{Easley2012-kj,
  title     = "Networks, crowds, and markets: Reasoning about a highly connected world",
  author    = "Easley, David and Kleinberg, Jon",
  publisher = "Cambridge University Press",
  year      =  2012
}

@INPROCEEDINGS{luoyiqian,
  author={Luo, Bing and Li, Xiang and Wang, Shiqiang and Huang, Jianwei and Tassiulas, Leandros},
  booktitle={IEEE Conference on Computer Communications}, 
  title={Cost-Effective Federated Learning Design}, 
  year={2021},
  volume={},
  number={},
  pages={1-10}
}

@INPROCEEDINGS{9488705,
  author={Tang, Ming and Wong, Vincent W.S.},
  booktitle={IEEE Conference on Computer Communications}, 
  title={An Incentive Mechanism for Cross-Silo Federated Learning: A Public Goods Perspective}, 
  year={2021},
  volume={},
  number={},
  pages={1-10}}

@MISC{StreamML,
  title        = "{StreamML}",
  booktitle    = "Stream.ml",
  howpublished = "\url{https://stream.ml/}",
  note         = "Accessed: 2025-9-14"
}

@MISC{GravityAI,
  title        = "{GravityAI}",
  booktitle    = "Gravity-ai.com",
  howpublished = "\url{https://www.gravity-ai.com/}",
  note         = "Accessed: 2025-12-18"
}

@MISC{Modelplace,
  title        = "Aimodelplace: Earn Money by selling your Artificial Intelligence Models",
  booktitle    = "Aimodelplace.com",
  howpublished = "\url{https://aimodelplace.com/}",
  note         = "Accessed: 2025-12-18"
}

@INPROCEEDINGS{Yan22,
  author={Yan, Yuping and Ligeti, Péter},
  booktitle={IEEE Conference on Information Technology and Data Science}, 
  title={A Survey of Personalized and Incentive Mechanisms for Federated Learning}, 
  year={2022},
  volume={},
  number={},
  pages={324-329}}

@InProceedings{pmlr-v119-karimireddy20a,
  title = 	 {{SCAFFOLD}: Stochastic Controlled Averaging for Federated Learning},
  author =       {Karimireddy, Sai Praneeth and Kale, Satyen and Mohri, Mehryar and Reddi, Sashank and Stich, Sebastian and Suresh, Ananda Theertha},
  booktitle = 	 {International Conference on Machine Learning},
  pages = 	 {5132--5143},
  year = 	 {2020}
}

@inproceedings{NIPS2017_6211080f,
 author = {Smith, Virginia and Chiang, Chao-Kai and Sanjabi, Maziar and Talwalkar, Ameet S},
 booktitle = {Advances in Neural Information Processing Systems},
 pages = {1-11},
 title = {Federated Multi-Task Learning},
 year = {2017}
}

@ARTICLE{liang2020think,
      title={Think Locally, Act Globally: Federated Learning with Local and Global Representations}, 
      author={Paul Pu Liang and Terrance Liu and Liu Ziyin and Nicholas B. Allen and Randy P. Auerbach and David Brent and Ruslan Salakhutdinov and Louis-Philippe Morency},
      year={2020},
      journal={arXiv preprint arXiv:2001.01523},
      primaryClass={cs.LG}
}

@ARTICLE{Trevor22,
author = {Trevor Hastie and Andrea Montanari and Saharon Rosset and Ryan J. Tibshirani},
title = {{Surprises in high-dimensional ridgeless least squares interpolation}},
volume = {50},
journal = {The Annals of Statistics},
number = {2},
pages = {949 -- 986},
year = {2022}
}

@article{Mei22,
author = {Mei, Song and Montanari, Andrea},
title = {The Generalization Error of Random Features Regression: Precise Asymptotics and the Double Descent Curve},
journal = {Communications on Pure and Applied Mathematics},
volume = {75},
number = {4},
pages = {667-766},
year = {2022}
}

@article{AKrogh_1992,
year = {1992},
publisher = {},
volume = {25},
number = {5},
pages = {1135-1147},
author = {A Krogh and  J A Hertz},
title = {Generalization in a linear perceptron in the presence of noise},
journal = {Journal of Physics A: Mathematical and General}
}

@BOOK{mas1995microeconomic,
  title     = "Microeconomic Theory",
  author    = "Mas-Colell, Andreu and Whinston, Michael D and Green, Jerry R",
  publisher = "Oxford University Press",
  year      =  1995
}

@ARTICLE{10476711,
  author={Liao, Guocheng and Luo, Bing and Feng, Yutong and Zhang, Meng and Chen, Xu},
  journal={IEEE Transactions on Mobile Computing}, 
  title={Optimal Mechanism Design for Heterogeneous Client Sampling in Federated Learning}, 
  year={2024},
  volume={23},
  number={11},
  pages={10598-10609}}

@ARTICLE{hu2023,
      title={Federated Learning as a Network Effects Game}, 
      author={Shengyuan Hu and Dung Daniel Ngo and Shuran Zheng and Virginia Smith and Zhiwei Steven Wu},
      year={2023},
      journal={arXiv preprint arXiv:2302.08533},
      primaryClass={cs.LG},
}

@ARTICLE{karimireddy2022mechanisms,
      title={Mechanisms that Incentivize Data Sharing in Federated Learning}, 
      author={Sai Praneeth Karimireddy and Wenshuo Guo and Michael I. Jordan},
      year={2022},
      journal={arXiv preprint arXiv:2207.04557},
      primaryClass={cs.GT}
}

@INPROCEEDINGS{Chen2019-js,
  title     = "Demonstration of nimbus: Model-based pricing for machine
               learning in a data marketplace",
  booktitle = "International Conference on Management
               of Data",
  author    = "Chen, Lingjiao and Wang, Hongyi and Chen, Leshang and Koutris,
               Paraschos and Kumar, Arun",
    pages = {1885-1888},
  year      =  2019
}

@ARTICLE{10.14778/3447689.3447700,
author = {Liu, Jinfei and Lou, Jian and Liu, Junxu and Xiong, Li and Pei, Jian and Sun, Jimeng},
title = {Dealer: An End-to-End Model Marketplace with Differential Privacy},
year = {2021},
volume = {14},
number = {6},
journal = {VLDB Endowment},
pages = {957-969},
numpages = {13}
}

@inproceedings{10.1145/3328526.3329589,
author = {Agarwal, Anish and Dahleh, Munther and Sarkar, Tuhin},
title = {A Marketplace for Data: An Algorithmic Solution},
year = {2019},
booktitle = {ACM Conference on Economics and Computation},
pages = {701-726},
numpages = {26}
}

@ARTICLE{10571602,
  author={Huang, Wenke and Ye, Mang and Shi, Zekun and Wan, Guancheng and Li, He and Du, Bo and Yang, Qiang},
  journal={IEEE Transactions on Pattern Analysis and Machine Intelligence}, 
  title={Federated Learning for Generalization, Robustness, Fairness: A Survey and Benchmark}, 
  year={2024},
  volume={46},
  number={12},
  pages={9387-9406}}

@MISC{Online_appendix,
title = "Supplementary Materials: Mechanism Design for Federated Learning with Non-Monotonic Network Effects",
author="Li, Xiang and Luo, Bing and Huang, Jianwei and Luo, Yuan",
howpublished = "\url{https://drive.google.com/drive/folders/1Ktfj6cBKjKUNA0DrNyWqOztFYS3b2ARc?usp=sharing}",
note         = "Accessed: 2025-12-18"
}

@ARTICLE{Fernandez2020-cx,
  title    = "Data market platforms: Trading data assets to solve data problems",
  author   = "Fernandez, Raul Castro and Subramaniam, Pranav and Franklin,
              Michael J",
  journal  = "VLDB Endowment",
  volume   =  13,
  number   =  12,
  pages    = "1933--1947",
  year     =  2020
}

@inproceedings{10.1145/3299869.3300078,
author = {Chen, Lingjiao and Koutris, Paraschos and Kumar, Arun},
title = {Towards Model-Based Pricing for Machine Learning in a Data Marketplace},
year = {2019},
booktitle = {International Conference on Management of Data},
pages = {1535 - 1552},
numpages = {18}
}

@ARTICLE{Cong2022-pp,
  title    = "Data pricing in machine learning pipelines",
  author   = "Cong, Zicun and Luo, Xuan and Pei, Jian and Zhu, Feida and Zhang,
              Yong",
  journal  = "Knowledge and Information Systems",
  volume   =  64,
  number   =  6,
  pages    = "1417--1455",
  year     =  2022
}

@ARTICLE{hestness2017deep,
  title  = "Deep Learning Scaling is Predictable, Empirically",
  author = "Hestness, Joel and Narang, Sharan and Ardalani, Newsha and Diamos,
            Gregory and Jun, Heewoo and Kianinejad, Hassan and Patwary, Md
            Mostofa Ali and Yang, Yang and Zhou, Yanqi",
  year   =  2017,
  journal={arXiv preprint arXiv:1712.00409},
  primaryClass={cs.LG}
}

@inproceedings{
wang2024a,
title={A Lightweight Method for Tackling Unknown Participation Statistics in Federated Averaging},
author={Shiqiang Wang and Mingyue Ji},
booktitle={The Twelfth International Conference on Learning Representations},
year={2024},
pages={1-12}
}

@ARTICLE{Lecun,
  author={Lecun, Y. and Bottou, L. and Bengio, Y. and Haffner, P.},
  journal={Proceedings of the IEEE}, 
  title={Gradient-based learning applied to document recognition}, 
  year={1998},
  volume={86},
  number={11},
  pages={2278-2324}}

@ARTICLE{krizhevsky2009learning,
  title={Learning multiple layers of features from tiny images},
  author={Krizhevsky, Alex and Hinton, Geoffrey and others},
  year={2009},
  pages = {32-33},
  journal={Toronto, ON, Canada}
}

@inproceedings{37648,
title	= {Reading Digits in Natural Images with Unsupervised Feature Learning},
author	= {Yuval Netzer and Tao Wang and Adam Coates and Alessandro Bissacco and Bo Wu and Andrew Y. Ng},
year	= {2011},
booktitle	= {NIPS Workshop on Deep Learning and Unsupervised Feature Learning},
pages = {1-9}}

@inproceedings{
Li2020On,
title={On the Convergence of FedAvg on Non-IID Data},
author={Xiang Li and Kaixuan Huang and Wenhao Yang and Shusen Wang and Zhihua Zhang},
booktitle={International Conference on Learning Representations},
year={2020},
pages = {1-26}
}

@inproceedings{hu2023generalization,
  title={Generalization bounds for federated learning: Fast rates, unparticipating clients and unbounded losses},
  author={Hu, Xiaolin and Li, Shaojie and Liu, Yong},
  booktitle={International Conference on Learning Representations},
  year={2023},
  pages = {1-41}
}


 





\end{document}